\documentstyle[amsfonts,amssymb,preprint,aps]{revtex}
\tightenlines
\begin{document}

\title{Graph kinematics of discrete physical objects:\\
beyond space-time. III. Heisenberg --- Dyson's\\
two-layer physics approach}
\author{V. E. Asribekov}
\address{All - Russian Institute for Scientific and
Technical Information, VINITI, Moscow 125315, Russia\\
({\rm e-mail:peisv@.viniti.ru})}
\maketitle

\begin{abstract}
In part III is realized the consistent development of
Heisenberg---Dyson's two-layer matrix approximation to the graph
formalism for postulating discrete physical objects (DPO)
introduced in parts I--II in the form of discrete sets of
graphs---skeleton ({\bf SvT}) or root ({\bf RvT}) $v$-trees,
beyond common space---time. It is noted that already in the
late-1950s one made an attempt to formulate in physical theory the
discontinuity as an element of some special diagram technique. In
the framework of pointed Heisenberg---Dyson's two-layer matrix
scheme, with an incidence {\bf I} and a loop {\bf CD}$(\delta)$
graph matrices, are got the following main results: (1) the
many-``planes'' {\bf SvT} or {\bf RvT} representation of any DPO
in opposition to one-``plane'' physical objects in continuous
physical models; (2) the superposition of different types of
interaction for any microobject where {\bf RvT} representation for
short-ranged interactions (weak, strong) is one-``plane'' and for
long-ranged interactions (gravitational, electromagnetic) is
many-``planes''; (3) based on the incidence matrix {\bf I} (upper
layer) ``graph geometry'' of real DPO describes their peculiar
many-``planes'' inner structure beyond common space---time; (4)
the notion of interacting ``charge'' can be extracted only from the
symbolical quantities for the quasi-continuous field ``objects''
by means of the loop matrix {\bf CD}($\alpha$) (under layer); (5)
the set up strong correlation between upper (discrete material
objects) and under (quasi-continuous field ``objects'') layers of
full two-layer matrix, with the help of Maxwell's equations, may
be eliminated in the frame of DPO conception, by the way of
transition to the ``universal linearity'' beyond space---time; (6)
the problems of a stability (reproduction) of stationary states, of
a transition itself (``jump'') of electron and of a sudden
emission (``production'') of photon in atoms are solved directly
for stable atomic shell system in many-``planes'' {\bf RvT}
representation including its permanent reproduction with
non-open ``transferability'' between ``graph isomers''; (7) the
possibility of a ``fatigue'' of atomic shell system is discussed;
(8) the introduction of ``true discontinuity'' for structural
discrete microobjects (instead of ``QN-discontinuity'' from
quantum mechanics) and the realization of the ``top hierarchy''
with a superposition sum of Riemann's ``counting homogeneous
elements'' for {\bf RvT}s of constituent parts and a creation of
($r,s$) --- subsequences are carried out; and some other concrete
results of an analysis of structural peculiarities of DPO.
\end{abstract}

PACS: 11.90, 12.90, 02.10

\section*{1 Introduction}

According to Dyson's opinion in well-known paper (Ref. [1]) ---
see also Ref. [2] --- the correct fundamental theories will be
developing only after twentieth century.

\subsection*{1.1 Heisenberg---Dyson's analysis of Maxwell theory}

At the same time Dyson (Ref. [1]) as well as Heisenberg (Ref. [3])
were set up that in the broadest sense a central idea of Maxwell
theory, how it is considering now, can be formulated as a
conception of the two-layer structure of nature. In the under more
deep layer we have the electromagnetic fields, peculiar ``objects''
with simple wave equation's  description, and in the upper layer we
can observe the real material objects, their energy and forces. The
symbolical field quantities for so-called ``objects'' from an
under layer such as electric and magnetic field strengths, the main
continuous quantities in Maxwell's equations, are the purely
mathematical abstractions (for a given case see particularly the
special Weyl's investigation in Ref. [4]) and for this reason can
be determined only through their energy and forces in an upper
layer.

In general these results are true for the same continuous
quantities of any other fields that on the whole give rise to an
unsound situation which will alter completely only by the
transition to a new mathematical description of the basic real
physical objects taking into account a possible their discontinuity
and a probable changing of the existed correlation between under
and upper layers (see especially the foundation papers of Maxwell,
Einstein and Bohr in Refs. [5--7]).

\subsection*{1.2 On the new fundamental theories}

Indeed therefore one can consider the possibility of any concrete
realization of the indicated perspective Heisenberg---Dyson's
conception of two-layer physics (see Refs. [1, 3]) with new
adequate mathematical description of the basic real physical
objects, including both discrete (predominantly) and continuous
ones, as a steppingstone toward the cardinal changes of competing
contemporary fundamental theories --- the quantum field theory, the
$S$-matrix theory, the group theory (see, for example, Ref.
[8]) --- and also toward the most natural propositions on the
initial principles for future fundamental theories, all the more
that an introduction of some accepted in the present-day physics
principles was done to a great extent ``by hands''. From these new
principles above all one must call --- besides {\it a
discontinuity} --- {\it a hierarchy} (a subjection and a
consubjection), {\it a reproduction} (an immediate or an
intermediate mechanism), {\it an integrity of discrete physical
object and its non-reducing as a whole to the sum of parts} (a
correlation of the whole and its quasi-autonomic parts), {\it a
replication} (an evolution of structures with one or many
hierarchical centres), {\it a non-overcumulation} (particularly,
in graph vertices, analogously to an equation of continuity), and
so on, as distinct from earlier evident primary principles of {\it
an inertia}, {\it a stability}, {\it a space --- time extension},
{\it a superposition}, etc., and also derivative principles of
{\it an equilibrium},  {\it a homogeneity and an isotropy of any
space},  {\it a symmetry}, etc.

Once more it is necessary to emphasize that the creating
fundamental theories with ordinary continuous physical objects and
with postulated discrete (or quasi-continuous) physical objects
are the antipodes which mutually cannot serve as an justification
of one for another that is pointed out yet in Introduction
(section~1) to part~I of paper and especially in Introduction
(subsection 1.1) to part~II of paper, in agreement with the
Riemann's postulates. As it is noted by Dirac (see Ref. [9]):
``there will have to be some new development that is quite
unexpected, that we cannot make a guess about, which will take us
still further from classical ideas\ldots{} But if we cannot find
such a way \ldots{} we simply have to take into account that we
are at a transitional stage and that perhaps it is quite
impossible to get a satisfactory picture for this stage''.

\subsection*{1.3 Graph kinematics for postulating discrete physical \\
objects beyond space~--- time}

Let us return to parts I--II of paper (see Refs. [10--11]) where it
is described a graph kinematics of some postulating discrete
physical objects as the one of possible ways to realization of such
two-layer physics conception. For a full physical process graph,
including subgraphs of separate fragments of the structural physical
objects, is introduced the universal, beyond space---time,
formalism based on the special two-layer matrix
$$\displaystyle{\bf M}(\delta)=\left\{{{\bf I}\atop{\bf
CD(\delta)}}\right\}, \eqno (1)$$

\noindent
where $\delta\equiv\omega,\alpha,\ldots,$ etc., composed from an
incidence {\bf I} and a loop {\bf CD}($\delta$) graph matrices.
This adequate mathematical formalism, in agreement with the
above-mentioned Heisenberg---Dyson's two-layer scheme (see Refs.
[1, 3]), allows to represent any real discrete physical object and
each derivative quasi-continuous field ``object''. Furthermore it
could be led to the concrete analysis of the basic notions and
quantities, in accordance with structural peculiarities of these
objects, and to the following their interpretation. In spite of the
limited number of non-oriented or oriented graph parameters,
namely $v$ and $n=v-1$ in particular case of non-directed or
directed $v$-trees, there are just enough possibilities for
determination of the different characteristics of everyone from the
pointed objects only by the way of various graph vertices
configurations, i. e. we obtain factually a specific graph method
for description of the real discrete physical objects including
their interactions in physical processes of all kinds.

Practically from parts I and II of paper we have already in the
framework of appropriate aspects of the graph theory for
non-directed as well as for directed trees technique (initiated by
Kirchhoff)

--- {\it a ``graph geometry'' for the real discrete
physical objects, described their structure by means of an
incidence matrix $\displaystyle \mathop{{\bf I}}_{{(v-1
\times n)}}$ (upper layer) with the proper skeleton or root trees
basis where $v$ is a number of vertices and $n$ is a number of
lines (``edges'') of corresponding non-oriented ($v,n$)~--- graph,

--- an ``extremal equilibrium'' condition for the symbolical
quantities of quasi-continuous field ``objects'' as complex
systems, described all-round by means of a loop matrix
$\displaystyle \mathop{\bf C}_{{(l\times n)}}\cdot\mathop{{\bf
D}}_{{(n\times n)}}(\delta)$ (under layer) corresponding to the
$l$ different independent loops of suitable oriented
($v,n$)~--- graph with the same proper skeleton or root trees
basis}.

Below, beginning from section 2, we will go on a development of
postulated earlier graph kinematics formalism.

A simplicity of the graph mathematics, caused apparently by the
rejection from the ``Extension learning'' (die Ausdehnungslehre) in
the frame of generally accepted notion of space---time and by the
transition to a new conception with the natural insertion of
so-called ``Binding learning'' (die Verbindungslehre) already
beyond space---time in old understanding, may be demonstrated with
the help of the essentially linear principles including the linear
principle of superposition in graph technique and the corresponding
additional linear principles of a ``non-overcumulation'' in graph
vertices and a ``cyclic stability'' along independent graph loops.

\subsection*{1.4 On a possible evolution of dimensional symmetry\\
in the physicist's picture of nature}

The transition from an old mathematical scheme to the graph
kinematics formalism beyond space---time does not exclude the
dimensions problem. It will emerge also in the following introduction of
the new fundamental physical theories.

However, according to Dirac's conclusion in due time (see Ref.
[9]), then obtained results within the framework of space---time
consideration had inevitably led us to doubt how fundamental the
four-dimensional requirement in physics is: ``A few decades ago
it seemed quite certain that one had to express the whole of
physics in four-dimensional form. But now it seems that
four-dimensional symmetry is not of such overriding importance,
since the description of nature sometimes gets simplified when one
departs from it''. In this connection, if already a picture of
nature with four dimensions, after the pass from a picture of
nature with three dimensions, is not completely symmetrical, one
can formulate the next very general propositions:

1.~it is apparently a false direction of development of theoretical
models with a progressive increase of dimensions as follows from an
above-noted not quite perfect four-dimensional symmetry (see Ref.
[9]),

2.~it is probably a non-adequate way of postulating of the
``fractals'' or finding of the more hidden types of dimensional
symmetry as a key physical notions for new theoretical models,
particularly in case of microphysics,

3.~finally, taking into account 1. and 2., one must simply
postulate an absence of common space---time, at least in
microworld.

Obviously this last conclusion 3. corresponds to an initiate
assumption of this paper.

It is important to note also that graph is topologically an
one-dimensional simplicial complex or a linear complex which
consists from vertices (points) and ``edges'' (lines) (see, for
example, Ref. [12]).

\pagebreak
\section*{2 MICROWORLD PHYSICS IN THE FRAME OF GRAPH\\
KINEMATICS FORMALISM}

The extraordinary complicated main subject of microworld physics
has been forming by quantum theory for the past almost 100 years.
However the essential successes in a solution of this immensely
hard problem of the quantum theory came only with the discovery of
quantum mechanics.

\subsection*{2.1 Dyson's analysis of quantum mechanics and
discrete\\
microobjects problem}

According to Dyson's conclusion (see Ref. [1]) a central idea of
quantum mechanics, again in the broadest sense, can be realized by
an extension to microworld of the two-layer structure conception
as it was occured already for Maxwell theory. At that the
proposition about two-layer construction of the nature in quantum
mechanics leads us, as a next step, to the general physical theory,
classical as well as quantum, which is more consistent, realistic
and universal itself than a former one.

In the under layer we have now the electric, magnetic and other
fields with their strengths and in addition the mathematical
abstractions of the same kind, namely the wave functions which
describe a behaviour of different microobjects. In the upper layer
besides the energy and forces of any real objects there appear the
probabilities of occurrence of various events.

Thus we come factually to an unified two-layer picture for
Maxwell's theory and quantum mechanics, at least roughly.
Nevertheless there are lots of distinctions in details between a
classical electromagnetism picture and a behaviour of microobject;
all the more the latter has an enough inexact or inaccurate
definition itself. Apparently all microobjects must be discrete and
therefore could be described generally by means of a special
mathematical formalism which is to follow from the specific
microworld conditions expressing the primary principles of
discontinuity.

On the other hand, in part I of paper (see Ref. [10]) it is assumed
that one can choose, in particular, the transition itself to
microworld physics as an optimum stage for postulating of the
discrete physical microobjects in a form of the new ``physical
graphs''. On this assumption one must use the old Feynman diagram
technique as a starting point in frame of the $S$-matrix theory,
which intense-investigated earlier (see Ref. [8] and references in
[10]) together with the field theory and the group theory.

Just in those years Landau (see Ref. [13, 2]) had fixed that the
specific diagram technique for determination of singularities in
the quantum field theory quantities is really beyond a formalism of
this theory. Taking into account a non-equivalence of such ``old''
diagram technique, based only on the stable microobjects~--- both
``simple'' and ``complex'', to a perturbation theory, Landau had
decided that solely a straight use of diagram technique is
completely consistent and under such estimation we should make
factually the first steps to successive development of a ``new''
diagram technique which is the generalization of an ``old'' one. It
is very likely that this ``new'' diagram technique itself could
serve, according to Landau, as a ground for an adequate future
fundamental theory. To a certain extent similar put forward
hypothesis one must take also into consideration at least as an
initiate attempt to do without the customary continuous physical
theory.

\subsection*{2.2 General survey of two-layer conception in
microphysics\\
and deterministic picture problem}

Owing to the set up uniformity of a possible realization of
two-layer conception in macrophysics (Maxwell's theory) as well as
in microphysics (quantum mechanics) we can carry out the following
typical specification of the graph kinematics formalism (see
subsection 1.3) first of all in an area of investigation of the
discrete microobjects (or their discrete parts) which appear
inevitably in this theoretical scheme. Without going into details
one can note that for every separate discrete microobject, from the
whole physical system of microobjects, an incidence matrix {\bf I}
in (1) gives as usually a representation of its inner structure (or
inner structure of its discrete parts) on the base of the
non-directed root trees
vertices configuration, while a loop matrix {\bf CD}($\alpha$) in
(1) reflects the loop-forming quasi-continuous field ``object''
produced by a ``fusion'' of several discrete parts of microobjects
together or, more exactly, by a corresponding superposition of
their directed root trees vertices with conservation of the total
number of lines (``edges''). For example, in a case of the most
evident forecast of ``fusion'' or vertex-corresponding
superposition  of two leptons {\it e}$^{-}$ and $e^{+}$ as the
discrete microobjects, having an elementary form of two-side
directed root ({\it v}=2)-trees:

\vskip8dd
\rule{50mm}{0dd}
\begin{picture}(33,3)
\put(0,0){\vector(1,0){30}}
\put(0,0){\circle*{3}}
\put(0,0){\circle{6}}
\put(30,0){\circle*{3}}
\end{picture}
\raisebox{-2pt}{$(e^{-})$ \ \rm and}
\rule{10dd}{0dd}
\begin{picture}(33,3)
\put(30,0){\vector(-1,0){30}}
\put(0,0){\circle*{3}}
\put(0,0){\circle{6}}
\put(30,0){\circle*{3}}
\end{picture}
\raisebox{-2pt}{$(e^{+})$},
\ \hfill\raisebox{-2pt}{($\ 2a$)}
\vskip8dd

\noindent
one may create a photon
h$\nu$ or the simplest quasi-continuous field ``object'', having
a form of oriented multigraph---semicycle:

\vskip8dd
\rule{47mm}{0dd}
\begin{picture}(40,16)
\put(0,0){\line(0,1){16}}
\put(0,8){\circle*{3}}
\put(0,8){\circle{6}}
\put(0,16){\line(1,0){40}}
\put(12,16){\vector(1,0){10}}
\put(0,0){\line(1,0){40}}
\put(27,0){\vector(-1,0){10}}
\put(40,0){\line(0,1){16}}
\put(40,6){\circle*{3}}
\end{picture}
\raisebox{3pt}{\rule{3pt}{0dd}$(h\nu)\ {\rm with}\  v=2, \
l=1.$} \ \hfill\raisebox{3pt}{$(\ 2b$)} \vskip8dd

In general, it is significant that a transformation of non-oriented
graphs into oriented ones within the framework of a developing
graph formalism is carried out with the help of a doubling~---
operation consisted in a substitution of every or separate
non-directed line (``edge'') by a cycle pair of opposite directed
lines (``edges'') (see, for example, Ch. 1 in [19]), analogous to
above---represented in (2b) oriented multigraph~--- semicycle of
photon.

Additionally, in connection with obtained results it will be noted
that an ordinary description of microobjects in contemporary
general quantum theory of fields and particles by means of the
probability distributions for continuous quantities in a
space---time become apparently insufficient in a case of the graph
kinematics scheme beyond space---time, with a reliable direct
``count''  for evident discrete quantities (homogeneous elements)
in elementary physical processes. Perhaps the mere fact that this
giving up of determinacy has been a very controversial subject,
which is due certainly to quantum theory and some physicists
(Einstein especially) do not like it at all, shows simply a
non-adequacy of the picture of nature in the infinitesimal to the
space---time basis notions.

In some or  other way, a today's ``continuous'' picture of
microworld with probability description of microobjects in a
space---time, which had always taken for granted, will be changed
radically under the more obvious ways of a transition to the graph
kinematics formalism beyond space---time for the doubtless discrete
microobjects.

\pagebreak
\subsection*{2.3 Basic graph characteristics for microobjects\\
and Ulam's hypothesis.}

Awaring the importance of studying of the main properties of
graphs, and especially trees, which could serve as a basic tool for
description of microobjects within the graph kinematics, we pass to
the consideration of a set of concrete graph characteristics for
microobjects and special hypotheses from the graph theory.

Beforehand one must introduce for graph {\bf G} with {\it v}
vertices the {\it power} of vertex $v_{i}$
$$\displaystyle {\frak M}(v_{i})\equiv {\frak M}_{i} \eqno (3)$$

\noindent
i. e. a number of components for graph {\bf G}$_{i}$={\bf G}--{\bf
v}$_{i}$. According to Ulam's hypothesis in  Harary's interpretation
(see Ref. [12]) for a case $ v \geqslant $3 one takes place the
reconstruction (or, more exactly, the reproduction) of a graph {\bf
G} from a collection of subgraphs {\bf G}$_{i}$={\bf G}---{\bf
v}$_{i}$. The proof of Ulam's hypothesis for trees had presented by
Kelly (see Ref. [14]). But later the Kelly's result had generalized
by Harary and Palmer (see Ref. [15]) and at last had improved by
Bondy (see Ref. [16]). Starting from propositions of Ulam's
hypothesis, these authors had proved a possibility of the
reconstruction (the reproduction) of a tree {\bf T} from a
collection of subtrees {\bf T}$_{i}$={\bf T}---{\bf
v}$_{i}$ where $v_{i}$ are the terminal (Harary, Palmer) or
the peripheral (Bondy) vertices; for the latter an eccentricity
$$\displaystyle e_{ T}(v_{i})=d_{\rm max} (u, v_{i}), \eqno
(4a)$$

\noindent
where {\it u}~--- any vertex of {\bf T}, is equal to a diameter of
{\bf T}
$$\displaystyle D_{T}=\max\limits_{(v_{i})} e_{T} (v_{i}); \
i=1,2,\ldots{} , v. \eqno (4b)$$

\noindent
These convenient formulations of Ulam's hypothesis for trees can be
put forward as the initial theorems for a graph consideration of
stable and quasi---stable physical systems in microworld.
Additionally it will be noted also that for trees the power of
vertex $v_{\rm i}$ is equal to its degree
$$\displaystyle {\frak M}(v_{i})={\bf deg} v_{i}. \eqno (5)$$

Taking into account that we separate from the all trees only the
root {\it v}-trees ({\bf RvT}) with one detached root vertex
({\it v}$_{R}$) from a total number of vertices {\it v} (see
section 3 in part II, Ref. [11]) one can thereby determine the
most ``chief'' {\bf RvT} which includes only ``core'' vertices
($v_{ R}$) and ($v^{c}_{F}$) with $v_{ R}$=1, $v_{ F}^{\rm
c}=v-1$, $v=v_{ R}+v_{ F}=1+v^{ c}_{ F}, \ d_{\rm
min}(v_{R}, v_{F})\equiv d(v_{R}, v^{c}_{F})=1$ and characterizes
the pure ``core'' of discrete microobject. Here solely one vertex
($v_{R}$) possesses the maximal power ${\frak M}_{R}\equiv {\frak
M}_{\rm max}(v_{R})=v-1$  whereas the other {\it v}~---1 vertices
($v^{c}_{F}$) of ``core'' have the minimal power ${\frak M}_{\rm
min}(v^{c}_{F})$=1. At first sight it is seemed that, owing to
maximal value of the power ${\frak M}_{R}$ for a single root vertex
($v_{R}$) in the pure ``core'', just this ``chief'' {\bf RvT}
reffects the principal inner structure of corresponding discrete
microobject, as it were its ``genetics'', and the properties of its
various interactions with the another microobjects. Probably the
rest {\bf RvT}s represent mainly the peculiarities of a
constitution of different microworld systems and their typical
reactions.

\subsection*{2.4 About four sorts of ``charge'' in old concept}

At last going over to an interpretation of the main microobject's
graph characteristics, corresponding in the first place to
``charges'', it is important to emphasize that an appearance of
these notions in a new graph formalism may be connected with some
difficulties, especially if we follow a logic of independent
introduction of the primary oriented graph notions.

First of all, with a view to avoid any misunderstanding, we pay
attention again to an above---concerned problem of a setting up of
the characteristics of symbolical quantities for a
quasi-continuous field ``objects'', in a form of some
multigraphs, in comparison with the ones for continuous field. In
the framework of an old concept, a continuous field itself acquired
physical reality in a space---time, and an interaction of
microobjects had to be described usually with the help of an
intermediate, conditionally real, ``field of force''. It turned out
that the properties of microobjects with respect to an exchange
interaction with similar, conditionally real, ``field of force''
had to be determined by a single parameter~--- so-called the
specific ``charge'' of microobjects, which initially must be
different for the case of various ``fields of force''.

In very such a way one set up the four known fundamental
interactions whose distinctions had to be explained by multiple
exchange mechanisms as well as by intensity of these exchanges
(proportional to corresponding ``charges'') into a
``microobject~--- field'' system in conventional space---time, or
more exactly, in the presence of common metric characteristics of
space---time.

On the other hand, just at the like approach it is easy to come to
the conclusion that the most specific peculiarity of an old concept
concerning the fact of an existence only four possible sorts of
``charge'' (that is also in a formal agreement with the conclusion
of Hurwitz's theorem about four basic kinds of hypercomplex numbers
(four extraordinary algebras~--- see Ref.~[17])) may be
inconsistent with the proposals, immanent construction and general
principles of graph kinematics formalism. It is necessary to bear
in mind that an inner structure of microobject has to be included
into an incidence matrix $\displaystyle
\mathop{{\bf I}}_{(v-1\times n)}$ that is realized as
a ``graph geometry'' for the discrete microobject. However the
usual ``field's characteristics of interaction'' one must determine
with the help of a loop matrix {\bf CD}($\alpha$) i.~e. a
``charge'' or an interaction ``charge'' has to be extracted from the
new loop-forming quasi-continuous field ``objects''.

\subsection*{2.5 What is now a ``charge'' for?}

Indeed, due to the absence of the main notions of an old concept
and in exchange for a continuous, conditionally real, field and an
interaction of microobjects with corresponding intermediate
``field of force'' in space---time, these new field ``objects'' can
be represented beyond space---time as the oriented multigraphs with
some loops which are composed from two or more directed {\bf RvT}s.
At that in a given case we have {\it no parameters of interaction
for microobjects} but there are the original ``fusion'' or
vertex---corresponding superposition processes already for vertices
of the initial directed {\bf RvT}s and the resulting {\it oriented
multigraphs from which one can extract a notion of the graph
equivalent of something having a likeness with the ``charge''}.

\subsubsection*{2.5.1 Symbolical quantities for
loop-forming quasi-continuous \\
field ``objects'' as oriented multigraphs}

In developing formalism every microobject may be represented
integrally as a set of {\bf RvT}s i.~e. as a whole which consists
in a hierarchical order from many quasi-autonomous parts~---
independent ``planes'' generating a complicated ``graph geometry''
of structural microobject. Of course, an appearance of other
physical characteristics of this microobject may be provided only
for special symbolical quantities following from a consideration of
corresponding quasi-continuous field ``object''.

If one can write down the various symbolical quantities for the
loop-forming quasi-continuous field ``objects'' in a form
depending from different configurations of the oriented multigraph
vertices then the problem of an interaction ``charge''
interpretation will be reduced to an analysis of definite
procedures, including a graph reproduction of discrete
microobjects, a recombination of ``core''-vertices or any
reconstruction of corresponding microobject graphs (non-directed
and directed {\bf RvT}s and oriented multigraphs) using
particularly the results of an application of Ulam's hypothesis.
Later we will continue this graph approach by investigation of some
concrete examples.

As above, the symbolical quantities for quasi-continuous field
``objects'' are derived from an ``extremal equilibrium'' condition
with the help of a loop matrix {\bf CD}($\alpha$). In addition, it
is notable however that there exists a ``mechanical'' version of
interpretation of an analogous ``extremal conditions'' along
$l=n-v+1$ loops for an old Feynman diagram (see Ref.~[13]) which
earlier was discussed by Coleman and Norton (see Ref.~[18]). These
authors were shown that $l$ ``extremal conditions'' are equivalent
to the condition that the relevant Feynman diagram be interpretable
as a picture of an energy~--- and momentum~--- conserving process
occuring in space---time, with all internal microobjects real, on
the mass shell, and moving forward in time, for just the correct
distances and times to ``tie together'' the entire directed graph.
The Feynman parameters $\alpha _{i}$ ($i=1,\,2,\,\ldots{} \,n$),
associated with $n$ internal lines of this graph (see 4-momenta
$q_{i}$ in subsection 1.2 of part~I), were identified with the time
microobject exists between collisions, divided by its mass.

Thus, at a time taking into account this evident ``mechanical''
picture of the continuous microobjects interaction in space---time,
on the one hand, and assuming an ``invisible'' separate photon as
the elementary quasi-continuous field ``object'' (a simple
oriented multigraph in (2b) accounted for electromagnetic
interaction) beyond space---time, on the other hand, one can admit
also the another quasi-continuous field ``objects'' which are
represented as the various oriented multigraphs, reflected the
different interaction processes and the further procedures of decay
reconstruction and exchange reaction mechanisms, with the use of
initial discrete microobjects beyond space---time. Perhaps in such a
way we could solve the problem of a microobject ``charge''.

Nevertheless it will be noted that instead of various types of
fields and interactions with different ``charges'' of microobjects
in conventional space---time, we have now an unified {\bf RvT}~---
description where the different quasi-continuous field
``objects'' may be distinguished beyond space---time solely by the
sets of vertices in oriented multigraphs (composed from directed
{\bf RvT}s), their configurations and corresponding powers. From
an old concept one must remember still that each vertex interaction
could occur as an instantaneous event in space---time.

\subsubsection*{2.5.2 Graph equivalent
of microobject ``charge'' \\
and decoding of symbolical quantities}

Returning to Heisenberg---Dyson's analysis of Maxwell theory (see
subsection 1.1) one can go on the decoding of the symbolical
quantities for the loop-forming quasi-continuous field
``objects'' but keeping in view also a possible existence of some
prototypes in an old concept for a sought graph equivalent. This
problem may be solved, in particular, with the help of a
preliminary consideration and an analysis of Dirac's assumption
(Ref.~[9]) about a picture of ``discrete (quantized) Faraday lines
of force'', each associated with an electric charge ($-e$ or
$+e$).  Starting from an old idea that the Faraday lines of force
with continuous distribution are a way of picturing electric fields
and going over to quantum theory, Dirac brought a kind of
discreteness into a classical picture and replaced it by just a
whole number of discrete lines of force with no lines of force
between them. These lines of force in the Faraday picture end where
there are charges (for lines of force extending to infinity, of
course, there is no charge) and we shall have an explanation of why
charges always occur in multiples of one elementary unit, $e$, and
why does one not have a continuous distribution of charge occuring
in nature.  Some of these lines of force, forming closed loops or
simply extending from minus infinity to infinity, will correspond
to waves; others will have ends with the charges. The breaking of a
line of force would be the picture for the creation of an electron
$e^{-}$ and a positron $e^{+}$. One can consider also a picture of
the discrete lines of force as ``strings'', and then the electron
in the picture is the end of a ``string''.

Dirac's picture of ``discrete Faraday lines of force'' (or,
perhaps, a picture of some ``strings''), demonstrated an existence
of the elementary units~--- constants~--- of discrete charges, in
general different, may be used in our case (see (2a)) for the
setting up, analogously, an evident graph equivalent for an
elementary ``charge'' (a simplest directed {\bf RvT}) as well as
for corresponding an elementary ``inertial mass'' (a simplest
non-directed {\bf RvT}) of microobject. In subsection 3.2 of
part~I we have already a graph definition of an ``inertial mass''
$m_{e}$ for an electron in a form of the simplest non-directed
{\bf RvT} with $v$=2 ($T_{v=2}$=1). Thus one may postulate the
first ($v=2$, $n=1$)~--- graph definitions for ``charges'' $e^{-}$
and $e^{+}$ and ``inertial mass'' $m_{e}$

\vskip8dd
\rule{30mm}{0dd}
\begin{picture}(33,3)
\put(0,0){\vector(1,0){30}}
\put(0,0){\circle*{3}}
\put(0,0){\circle{6}}
\put(30,0){\circle*{3}}
\end{picture}
\raisebox{-2pt}{$(e^{-})$}
\rule{15dd}{0dd}
\begin{picture}(33,3)
\put(30,0){\vector(-1,0){30}}
\put(0,0){\circle*{3}}
\put(0,0){\circle{6}}
\put(30,0){\circle*{3}}
\end{picture}
\raisebox{-2pt}{$(e^{+})$}
\rule{15dd}{0dd}
\begin{picture}(33,3)
\put(30,0){\line(-1,0){30}}
\put(0,0){\circle*{3}}
\put(0,0){\circle{6}}
\put(30,0){\circle*{3}}
\end{picture}
\raisebox{-2pt}{$(m_{e})$}
\ \hfill\raisebox{-2pt}{($\,6\,$)}
\vskip8dd

\noindent
Naturally, the ``gravitational mass'' or scalar ``gravitational
charge'' has a graph equivalent also in a form of the simplest
directed {\bf RvT} as well as the other scalar ``charges'':
$g_{1}\equiv e$ (above---mentioned), $g_{2}\equiv G_{W}$,
$g_{3}\equiv g_{S}$.

Obviously, these introduced graph equivalents could contain also
the another characteristics of discrete microobject besides a
``charge'' and an ``inertial mass'', for example a ``spin'' (later
we revert to this problem).

Still for a clarity of an accepted approximation in direction of
the decoding of a meaning of every symbolical quantity which
describes the quasi-continuous field ``objects'' in the framework
of an oriented graph technique, one must take into consideration a
definite set of loop combinations~--- in accordance with a loop
matrix {\bf CD}($\alpha$) from (1)~--- of the different graph
equivalents, included in corresponding graph representation of the
various physical processes (decays, collisions, scattering,
reactions, etc.) as a rule with the conservation of a total number
of lines (``edges'') that is carried out by the use of the
vertex-corresponding superposition.

\subsection*{2.6 Superposition of different types of interaction}

The most unexpected feature of a developing new physical graph
technique is a possible {\it ``parallel'' presence of the four
(and more) different sorts of interaction ``charges'' all over for
any microobject} and for its symbolical quantities. All these
interaction ``charges'' have the unific graph representation in a
form of the various subsets from the full sets of vertices
($v_{i}$) in corresponding {\bf RvT}s for a given microobject.
Similar full sets of vertices ($v_{i}$) can be distinguished mainly
by their disposition towards a ``core'' of {\bf RvT} and, of
course, by their powers ${\frak M}$ ($v_{i}$), especially into a
``core'', and every interaction ``charge'' may occur in multiples
of one elementary ``unit'' presented by some number of {\bf RvT}'s
``core'' vertices of different powers. In other words, if we accept
such concept, the definite disposition of these vertices in a
``core'' subset and some combination of their powers must be
responsible for a concrete interaction ``charge''.

On the other hand, taking into consideration that certain physical
processes would be arisen out of an existence of fixed interaction
``charges'', one can specify in such a way the different
interaction ``charges'' of a single microobject. Thus we have a
criterion of identification of the different interaction ``charges''
of microobject on the base of a typical classification of various
elementary physical processes. It is easy to see also that one
cannot study the ``charges'' at rest as well as the corresponding
forces between them and the fields associated with them; in other
words we haven't a ``charge'' statics but there is only a so-called
``charge'' dynamics.

Many-``planes'' representation of any microobject as a set of
non-directed or directed {\bf RvT}s with a superposition of the
different sorts of interaction ``charges'', for the first time
mentioned in subsection 2.5.1, arranges now in Tables~1--3 where
are introduced also some ``graph---vacancies'' like {\bf
C}$_{v_{ F}}$ for further filling at $v_{ F}$=2$\div$4 and an
additional physically stable graph---compositions from several
known microobjects.

First of all in Table~1 we repeat (see above (2a), (2b), (6)) a
collection of the graph equivalents of microobject's masses,
interaction ``charges'' and their annihilation in a form of the
symplest non-oriented and oriented ($v$=2, $n$=2)~--- graphs (with
$l_{\max}$=1, $n_{\max}=v-1+l_{\max}$=2) which are accompanied by
the diverse graph matrices (see Ref.~[19]).

\pagebreak
\begin{center}

{\bf Table~1. Graph equivalents of basic microobject
characteristics and suitable matrices of these (v=2,
n=2)~--- graphs}.

\end{center}

\vskip 3dd

{\tabcolsep=4dd
\begin{center}
\newcommand{\Sloppy}{\emergencystretch 3em \tolerance 9999
}
\let\sloppy=\Sloppy
\begin{tabular}{p{53mm}|p{31mm}p{22mm}p{22mm}p{22mm}}
\hline
\multicolumn{1}{r|}{}                                           &
\multicolumn{1}{|c}{}                                           &
\multicolumn{1}{c}{}                                            &
\multicolumn{1}{c}{}                                            &
\multicolumn{1}{c}{}                                     \\[-5dd]
\multicolumn{1}{r|}{$(v=2, n=2)$-Graphs}                        &
\multicolumn{1}{|c}{}                                           &
\multicolumn{1}{c}{}                                            &
\multicolumn{1}{c}{}                                            &
\multicolumn{1}{c}{}                                           \\
\multicolumn{1}{r|}{ }                                          &
\multicolumn{1}{|c}{}                                           &
\multicolumn{1}{c}{}                                            &
\multicolumn{1}{c}{}                                            &
\multicolumn{1}{c}{Oriented}                                   \\
\multicolumn{1}{l|}{}                                           &
\multicolumn{1}{|c}{Non-Directed}                               &
\multicolumn{1}{c}{Directed}                                    &
\multicolumn{1}{c}{Directed}                                    &
\multicolumn{1}{c}{multigraph-}                                \\
\multicolumn{1}{l|}{}                                           &
\multicolumn{1}{|c}{\bf RvT     }                               &
\multicolumn{1}{c}{\bf RvT }                                    &
\multicolumn{1}{c}{\bf RvT }                                    &
\multicolumn{1}{c}{semicycle}                                  \\
\multicolumn{1}{l|}{Matrices}                                   &
\multicolumn{1}{|c}{\raisebox{-6pt}{\begin{picture}(52,30)%
\put(0,0){\line(1,0){50}}%
\put(0,0){\line(0,1){30}}%
\put(50,0){\line(0,1){30}}%
\put(0,30){\line(1,0){50}}%
\put(10,15){\line(1,0){30}}%
\put(10,15){\circle*{3}}%
\put(10,15){\circle{6}}%
\put(40,15){\circle*{3}}%
\end{picture}}%
\fbox{{\bf C}$_{v_{F}=1}$} }                                    &
\multicolumn{1}{c}{\raisebox{-6pt}{\begin{picture}(52,32)%
\put(10,15){\vector(1,0){30}}%
\put(10,15){\circle*{3}}%
\put(10,15){\circle{6}}%
\put(40,15){\circle*{3}}%
\end{picture}}}%
 &
\multicolumn{1}{c}{\raisebox{-6pt}{\begin{picture}(52,32)%
\put(40,15){\vector(-1,0){30}}%
\put(10,15){\circle*{3}}%
\put(10,15){\circle{6}}%
\put(40,15){\circle*{3}}%
\end{picture}}
}  &
\multicolumn{1}{c}{\raisebox{-6pt}{\begin{picture}(40,30)%
\put(0,0){\line(0,1){30}}%
\put(0,15){\circle*{3}}%
\put(0,15){\circle{6}}%
\put(0,30){\line(1,0){40}}%
\put(12,30){\vector(1,0){10}}%
\put(0,0){\line(1,0){40}}%
\put(27,0){\vector(-1,0){10}}%
\put(40,0){\line(0,1){30}}%
\put(40,15){\circle*{3}}%
\end{picture}
}}
                                                            \\[5dd]
\multicolumn{1}{l|}{of ($v$=2, $n$=2)-graphs}
& \multicolumn{1}{|c}{$(m_{e})$\rule{36dd}{0dd}   }             &
\multicolumn{1}{c}{$(e^{-})$}                                   &
\multicolumn{1}{c}{$(e^{+})$}                                   &
\multicolumn{1}{c}{(h$\nu$) }                                  \\
\multicolumn{1}{r|}{}                                           &
\multicolumn{1}{|c}{}                                           &
\multicolumn{1}{c}{}                                            &
\multicolumn{1}{c}{}                                            &
\multicolumn{1}{c}{}                                     \\[-5dd]
\hline
\multicolumn{1}{r|}{\rule{53mm}{0dd}}                         &
\multicolumn{1}{|c}{\rule{31mm}{0dd}}                         &
\multicolumn{1}{c}{\rule{22mm}{0dd}}                          &
\multicolumn{1}{c}{\rule{22mm}{0dd}}                          &
\multicolumn{1}{c}{\rule{22mm}{0dd}}                   \\[-5dd]
Vertex-edge\,incidence\,matrix                           &&&&\\
\multicolumn{1}{c|}{$\displaystyle \mathop{\bf I}_{(v-1\,\times n)}
(\epsilon)$ }                             &
\rule{0dd}{0dd}\hfill (1 \ 0) \hfill\rule{0dd}{0dd}         &
\rule{0dd}{0dd}\hfill (1 \ 0) \hfill\rule{0dd}{0dd}         &
\rule{0dd}{0dd}\hfill (0 \ --1) \hfill\rule{0dd}{0dd}        &
\rule{0dd}{0dd}\hfill (1 \ --1) \hfill\rule{0dd}{0dd}      \\
\multicolumn{1}{r|}{\rule{53mm}{0dd}}                         &
\multicolumn{1}{|c}{\rule{31mm}{0dd}}                         &
\multicolumn{1}{c}{\rule{22mm}{0dd}}                          &
\multicolumn{1}{c}{\rule{22mm}{0dd}}                          &
\multicolumn{1}{c}{\rule{22mm}{0dd}}                   \\[-5dd]
Vertex       incidence matrix      &&&&\\
\multicolumn{1}{c|}{$\displaystyle \mathop{\bf J}_{(v \times v)}
(\epsilon)$ }                                                   &
\multicolumn{1}{c}{
$\displaystyle \left({0 \atop 1} \
{1 \atop 0}\right)$  }                                          &
\multicolumn{1}{c}{
$\displaystyle \left({0 \atop -1} \
{1 \atop 0}\right)$   }                                         &
\multicolumn{1}{c}{
$\displaystyle \left({0 \atop 1} \
{-1 \atop 0}\right)$ }                                          &
\multicolumn{1}{c}{
$\displaystyle \left({0 \atop  0} \
{0 \atop 0}\right)$ }                                          \\
\multicolumn{1}{r|}{\rule{53mm}{0dd}}                         &
\multicolumn{1}{|c}{\rule{31mm}{0dd}}                         &
\multicolumn{1}{c}{\rule{22mm}{0dd}}                          &
\multicolumn{1}{c}{\rule{22mm}{0dd}}                          &
\multicolumn{1}{c}{\rule{22mm}{0dd}}                   \\[-5dd]
Edge         incidence matrix      &&&&\\
\multicolumn{1}{c|}{$\displaystyle \mathop{\bf K}_{(n \times n)}
(\epsilon)$ }                                                   &
\multicolumn{1}{c}{
$\displaystyle \left({0 \atop 0} \
{0 \atop 0}\right)$     }                                       &
\multicolumn{1}{c}{
$\displaystyle \left({0 \atop  0} \
{1 \atop 0}\right)$    }                                        &
\multicolumn{1}{c}{
$\displaystyle \left({0 \atop 1} \
{0  \atop 0}\right)$  }                                         &
\multicolumn{1}{c}{
$\displaystyle \left({0 \atop  1} \
{1 \atop 0}\right)$  }                                         \\
\multicolumn{1}{r|}{\rule{53mm}{0dd}}                         &
\multicolumn{1}{|c}{\rule{31mm}{0dd}}                         &
\multicolumn{1}{c}{\rule{22mm}{0dd}}                          &
\multicolumn{1}{c}{\rule{22mm}{0dd}}                          &
\multicolumn{1}{c}{\rule{22mm}{0dd}}                   \\[-5dd]
Loop  matrix    &&&&\\
$\displaystyle \mathop{\bf A}_{(l \times n)} (\alpha) =
\mathop{\bf C}_{(l \times n)}\cdot\,\mathop{\bf D}_{(n \times n)}
(\alpha)  $                                                    &
                                                                &
\multicolumn{1}{c}{
$\displaystyle (\alpha_{1}       \  0)$ }                      &
\multicolumn{1}{c}{
$\displaystyle (0 \ -\alpha_{2})$ }                            &
\multicolumn{1}{c}{
$\displaystyle (\alpha_{1} \ -\alpha_{2})$ }            \\
\end{tabular}
\end{center} }

\vskip 17dd

Here vertex incidence matrices $\displaystyle\mathop{{\bf J}}_{(2\times 2)}$
($\epsilon$) for $e^{-}$ and
$e^{+}$ may reflect a ``spin''---characteristics of these discrete
microobjects all the more that due to the absence of an imaginary
unit ``$i$'' beyond space---time for a reproduction of known
properties of Pauli matrices we have only:
$\displaystyle \mathop{{\bf J}}_{(2\times 2)}$ [$e
^{-}$]+$\displaystyle \mathop{{\bf J}}_{(2\times
2)}$ [$e^{+}$]=$\displaystyle \mathop{{\bf J}}_{(2\times
2)}$ [h$\nu$]; at the same time $\displaystyle \mathop{{\bf
J}}_{(2\times 2)}$ [$m_{e}$] may determine the other
``anti-charge'' (non-magnetic) intrinsic angular momentum of
$e^{\pm}$.  Simultaneously edge incidence matrix
$\displaystyle \mathop{{\bf
K}}_{(2\times 2)}$ ($\epsilon$) for
h$\nu$ must represent also some ``spin''-characteristic of photon
because a matrix equation:  $\displaystyle \mathop{{\bf
K}}_{(2\times 2)}$ [h$\nu$]=$\displaystyle \mathop{{\bf
K}}_{(2\times 2)}$ [$e^{-}$]+$\displaystyle \mathop{{\bf
K}}_{(2\times 2)}$ [$e^{+}$] can be considered analogously with
corresponding matrix equation for $\displaystyle \mathop{{\bf
J}}_{(2\times 2)}$ ($\epsilon$). In other words, both matrix
``spin''-characteristics for h$\nu$ appear as a summary
``spin''-characteristics of $e^{-}$ and $e^{+}$.

In the main from Table~1 follows that the loop-forming
quasi-continuous field ``object'' as an oriented ($v$=2,
$n$=2)---multigraph for h$\nu$ (semicycle) can be described now by
the respective simplest symbolical quantity with the help of loop
matrix {\bf A}($\alpha$)={\bf CD}($\alpha$)

$\displaystyle \mathop{{\bf A}}_{(1\times
2)}$($\alpha$)($\displaystyle \mathop{Q_{2}}_{(2\times
1)})$=($\alpha _{1}$--$\alpha _{2}$)$\displaystyle\left({q_{1}\atop
q_{2}}\right)$=$\alpha _{1}q_{1}-\alpha _{2}q_{2}$

what elucidates an use of evident
``mechanical'' version of interpretation of the ``extremal
equilibrium'' condition (see again a subsection 2.5.1) and allows
to set up a form of the ``source expression'' for an extraction of
the interaction ``charge'' notion (corresponding to the ``field's
characteristic of interaction'' in an old concept).

Shortly before a concrete analysis of the superposition of
different interactions on the basis of {\bf RvT}s at $v \geqslant 9$
for practically all subnuclear microobjects (because according to
Table~2 in part~II of paper (see Ref.~[11]) there is factually a
full spectrum of these microobjects at $v \geqslant 9$) we make
allowance also for some physically stable {\bf RvT}~---
compositions (quantum systems) and several ``graph---vacancies''
(for unknown physical microobjects) as an exception. Toward this end
in view we take into account at the first place a disposition and a
power of pure ``core'' vertices in corresponding {\bf RvT}s. At
best one may begin from the simplest pure ``core'', which is
characterized by minimum distance from root vertex ($v_{ R}$):
$d (v_{ R}, v_{ F}^{ c})=1$ for any
``free''---terminal---vertices ($v_{ F}^{\rm c}$), again in a
form of non-directed ($v=2$)~--- {\bf RvT} for ``$m_{\rm e}$''
(``core'' {\bf C}$_{v_{ F}=1}$) or directed ($v=2$)~--- {\bf
RvT} for ``e$^{\rm \pm}$'' indicated in Table~1. However it is more
interesting to consider in such a way the following {\bf RvT}s, all
over with a pure ``core'' separately, which correspond to an
interval $v=3 \div 8$ i.~e. to a region of atomic and nuclear
stable (shell) systems: see Table~1 in Part~II of paper
(Ref.~[11]). Below Table~2 contains in some detail only
the first part of a pointed interval, namely an interval $v=3\div5$
for the non-directed {\bf RvT}s.

\pagebreak
\begin{center}

{\bf Table~2. Non-directed ($\bf v, n=v-1$)~---{\bf RvT} for
$\bf 3\leqslant v\leqslant 5$ with summary power $\bf{\frak M}^{\bf
c}$ of ``core'' vertices}

\end{center}

\vskip 4dd

{\tabcolsep=3dd
\begin{center}
\newcommand{\Sloppy}{\emergencystretch 3em \tolerance 9999
}
\let\sloppy=\Sloppy
\begin{tabular}{p{15mm}|p{15mm}|p{43mm}p{43mm}|p{36mm}}
\hline
\multicolumn{1}{r|}{}                                           &
\multicolumn{1}{|c|}{}                                          &
\multicolumn{2}{|c|}{}                                          &
\multicolumn{1}{|c}{}                                    \\[-5dd]
\multicolumn{1}{c|}{Number}                                     &
\multicolumn{1}{|c|}{Number}                                    &
\multicolumn{2}{|c|}{{\bf RvT} --- compositions for stable
quantum}                                                        &
\multicolumn{1}{|c}{Pure ``core''  }                           \\
\multicolumn{1}{c|}{of    }                                     &
\multicolumn{1}{|c|}{of    }                                    &
\multicolumn{2}{|c|}{$(n_{QN}, \ l_{QN})$ --- systems
with   }                                                        &
\multicolumn{1}{|c}{{\bf RvT}\,s with }                        \\
\multicolumn{1}{c|}{vertices:}                                  &
\multicolumn{1}{|c|}{{\bf RvT}\,s:}                             &
\multicolumn{2}{|c|}{$3 \leqslant {\frak M}^{c} < 2
(v-1)$ }                                                        &
\multicolumn{1}{|c}{${\frak M}^{c}_{\rm max}=2(v-1)$}          \\
\multicolumn{1}{c|}{$v$      }                                  &
\multicolumn{1}{|c|}{$T_{v}$      }                             &
\multicolumn{2}{|c|}{}                                          &
\multicolumn{1}{|c}{as\,``graph-vacancies'' }                  \\
\multicolumn{1}{r|}{}                                           &
\multicolumn{1}{|c|}{}                                          &
\multicolumn{2}{|c|}{}                                          &
\multicolumn{1}{|c}{}                                    \\[-5dd]
\hline
\multicolumn{1}{r|}{\rule{15mm}{0dd}}                         &
\multicolumn{1}{|c|}{\rule{15mm}{0dd}}                        &
\multicolumn{1}{|c}{\rule{43mm}{0dd}}                         &
\multicolumn{1}{c|}{\rule{43mm}{0dd}}                         &
\multicolumn{1}{|c}{\rule{36mm}{0dd}}                  \\[-5dd]
&& \multicolumn{2}{|c|}{}                                       &
\multicolumn{1}{|r}{\fbox{${\bf C}_{v_{F}=2}$}}                \\
\multicolumn{1}{c|}{3}                                          &
\multicolumn{1}{|c|}{2}                                         &
\multicolumn{1}{|l}{\begin{picture}(73,30)%
\put(0,0){\line(1,0){50}}%
\put(0,0){\line(0,1){30}}%
\put(50,0){\line(0,1){30}}%
\put(0,30){\line(1,0){50}}%
\put(10,15){\line(1,0){60}}%
\put(10,15){\circle*{3}}%
\put(10,15){\circle{6}}%
\put(40,15){\circle*{3}}%
\put(70,15){\circle*{3}}%
\end{picture}%
\raisebox{12pt}{$l=0$}}                                      &
\multicolumn{1}{c|}{}                                           &
\multicolumn{1}{|l}{\rule{8dd}{0dd}\begin{picture}(50,40)%
\put(0,0){\line(1,0){50}}%
\put(0,0){\line(0,1){40}}%
\put(50,0){\line(0,1){40}}%
\put(0,40){\line(1,0){50}}%
\put(10,20){\circle*{3}}%
\put(10,20){\circle{6}}%
\put(10,20){\line(2,1){30}}%
\put(10,20){\line(2,-1){30}}%
\put(40,35){\circle*{3}}%
\put(40,5){\circle*{3}}%
\end{picture}%
}                  \\
\multicolumn{1}{c|}{ }                                          &
\multicolumn{1}{|c|}{ }                                         &
\multicolumn{1}{|l}{\rule{8dd}{0dd}${\frak M}^{c}=3$}         &
\multicolumn{1}{r|}{
\fbox{$\displaystyle l \equiv \mathop{l_{QN}=0\rule{12mm}{0dd}}^{
\displaystyle
n_{QN} = v - 2 = 1}$}}                                          &
\multicolumn{1}{|l}{\rule{8dd}{0dd}${\frak M}^{c}_{\rm
max}=4$}                                                       \\
\multicolumn{1}{r|}{\rule{15mm}{0dd}}                         &
\multicolumn{1}{|c|}{\rule{15mm}{0dd}}                        &
\multicolumn{1}{|c}{\rule{43mm}{0dd}}                         &
\multicolumn{1}{c|}{\rule{43mm}{0dd}}                         &
\multicolumn{1}{|c}{\rule{36mm}{0dd}}                  \\[-5dd]
\hline
\multicolumn{1}{r|}{\rule{15mm}{0dd}}                         &
\multicolumn{1}{|c|}{\rule{15mm}{0dd}}                        &
\multicolumn{1}{|c}{\rule{43mm}{0dd}}                         &
\multicolumn{1}{c|}{\rule{43mm}{0dd}}                         &
\multicolumn{1}{|c}{\rule{36mm}{0dd}}                  \\[-5dd]
\multicolumn{1}{c|}{ }                                          &
\multicolumn{1}{|c|}{ }                                         &
\multicolumn{1}{|r}{\begin{picture}(105,30)%
\put(0,0){\line(1,0){50}}%
\put(0,0){\line(0,1){30}}%
\put(50,0){\line(0,1){30}}%
\put(0,30){\line(1,0){50}}%
\put(10,15){\line(1,0){90}}%
\put(10,15){\circle*{3}}%
\put(10,15){\circle{6}}%
\put(40,15){\circle*{3}}%
\put(70,15){\circle*{3}}%
\put(100,15){\circle*{3}}%
\end{picture}%
\raisebox{12pt}{$l=0$}}                                      &
\multicolumn{1}{l|}{\begin{picture}(60,30)
\put(0,0){\line(1,0){50}}
\put(0,0){\line(0,1){30}}
\put(50,0){\line(0,1){30}}
\put(0,30){\line(1,0){50}}
\put(10,20){\circle*{3}}
\put(10,20){\circle{6}}
\put(10,20){\line(1,0){46}}
\put(33,20){\circle*{3}}
\put(56,20){\circle*{3}}
\put(10,20){\line(2,-1){23}}
\put(33,9){\circle*{3}}
\end{picture}\raisebox{12pt}{$l=1$}}                           &
\multicolumn{1}{|r}{\begin{picture}(54,40)
\put(0,0){\line(1,0){50}}
\put(0,0){\line(0,1){40}}
\put(50,0){\line(0,1){40}}
\put(0,40){\line(1,0){50}}
\put(10,20){\line(1,0){30}}
\put(10,20){\circle*{3}}
\put(10,20){\circle{6}}
\put(40,20){\circle*{3}}
\put(10,20){\line(2,1){30}}
\put(10,20){\line(2,-1){30}}
\put(40,35){\circle*{3}}
\put(40,5){\circle*{3}}
\end{picture}
\raisebox{32pt}{\fbox{${\bf C}_{v_{F}=3}$}}}                \\
\multicolumn{1}{r|}{\rule{15mm}{0dd}}                         &
\multicolumn{1}{|c|}{\rule{15mm}{0dd}}                        &
\multicolumn{1}{|c}{\rule{43mm}{0dd}}                         &
\multicolumn{1}{c|}{\rule{43mm}{0dd}}                         &
\multicolumn{1}{|c}{\rule{36mm}{0dd}}                  \\[-5dd]
\multicolumn{1}{c|}{4}                                          &
\multicolumn{1}{|c|}{4}                                         &
\multicolumn{1}{|l}{\rule{8dd}{0dd}${\frak M}^{c}=3$}         &
\multicolumn{1}{l|}{\rule{8dd}{0dd}${\frak M}^{c}=5$}         &
\multicolumn{1}{l}{\rule{8dd}{0dd}${\frak M}^{c}_{\rm max}=6$}
                                                                \\
\multicolumn{1}{r|}{\rule{15mm}{0dd}}                         &
\multicolumn{1}{|c|}{\rule{15mm}{0dd}}                        &
\multicolumn{1}{|c}{\rule{43mm}{0dd}}                         &
\multicolumn{1}{c|}{\rule{43mm}{0dd}}                         &
\multicolumn{1}{|c}{\rule{36mm}{0dd}}                        \\
\multicolumn{1}{c|}{ }                                          &
\multicolumn{1}{|c|}{ }                                         &
\multicolumn{1}{|l}{\begin{picture}(60,30)
\put(0,0){\line(1,0){50}}
\put(0,0){\line(0,1){30}}
\put(50,0){\line(0,1){30}}
\put(0,30){\line(1,0){50}}
\put(10,15){\line(1,0){23}}
\put(10,15){\circle*{3}}
\put(10,15){\circle{6}}
\put(33,15){\line(2,1){23}}
\put(33,15){\line(2,-1){23}}
\put(33,15){\circle*{3}}
\put(58,27){\circle*{3}}
\put(58,3){\circle*{3}}
\end{picture}
\rlap{\raisebox{24pt}{$l=1$}}{$l=1$}}                           &
\multicolumn{1}{l|}{ }                                          &
\multicolumn{1}{|l}{ }                                   \\[-5dd]
\multicolumn{1}{r|}{\rule{15mm}{0dd}}                         &
\multicolumn{1}{|c|}{\rule{15mm}{0dd}}                        &
\multicolumn{1}{|c}{\rule{43mm}{0dd}}                         &
\multicolumn{1}{c|}{\rule{43mm}{0dd}}                         &
\multicolumn{1}{|c}{\rule{36mm}{0dd}}                  \\[-5dd]
\multicolumn{1}{c|}{ }                                          &
\multicolumn{1}{|c|}{ }                                         &
\multicolumn{1}{|l}{\rule{8dd}{0dd}${\frak M}^{c}=4$}         &
\multicolumn{1}{r|}{
\fbox{$\displaystyle l \equiv
\mathop{l_{QN}=0, 1\rule{8mm}{0dd}}^{
\displaystyle
n_{QN} = v - 2 = 2}$}}                                          &
\multicolumn{1}{|l}{\rule{8dd}{0dd}}                         \\
\multicolumn{1}{r|}{\rule{15mm}{0dd}}                         &
\multicolumn{1}{|c|}{\rule{15mm}{0dd}}                        &
\multicolumn{1}{|c}{\rule{43mm}{0dd}}                         &
\multicolumn{1}{c|}{\rule{43mm}{0dd}}                         &
\multicolumn{1}{|c}{\rule{36mm}{0dd}}                  \\[5dd]
\end{tabular}
\end{center} }

\vskip 14dd

{\tabcolsep=3dd
\begin{center}
\newcommand{\Sloppy}{\emergencystretch 3em \tolerance 9999
}
\let\sloppy=\Sloppy
\begin{tabular}{p{15mm}|p{15mm}|p{43mm}p{43mm}|p{36mm}}
\hline
\multicolumn{1}{r|}{}                                           &
\multicolumn{1}{|c|}{}                                          &
\multicolumn{2}{|c|}{}                                          &
\multicolumn{1}{|c}{}                                    \\[-5dd]
\multicolumn{1}{c|}{Number}                                     &
\multicolumn{1}{|c|}{Number}                                    &
\multicolumn{2}{|c|}{{\bf RvT} --- compositions for stable
quantum}                                                        &
\multicolumn{1}{|c}{Pure ``core''  }                           \\
\multicolumn{1}{c|}{of    }                                     &
\multicolumn{1}{|c|}{of    }                                    &
\multicolumn{2}{|c|}{$(n_{QN}, \ l_{QN})$ --- systems
with   }                                                        &
\multicolumn{1}{|c}{{\bf RvT}\,s with }                        \\
\multicolumn{1}{c|}{vertices:}                                  &
\multicolumn{1}{|c|}{{\bf RvT}\,s:}                             &
\multicolumn{2}{|c|}{$3 \leqslant {\frak M}^{c} < 2
(v-1)$ }                                                        &
\multicolumn{1}{|c}{${\frak M}^{c}_{\rm max}=2(v-1)$}          \\
\multicolumn{1}{c|}{$v$      }                                  &
\multicolumn{1}{|c|}{$T_{v}$      }                             &
\multicolumn{2}{|c|}{}                                          &
\multicolumn{1}{|c}{as\,``graph-vacancies'' }                  \\
\multicolumn{1}{r|}{}                                           &
\multicolumn{1}{|c|}{}                                          &
\multicolumn{2}{|c|}{}                                          &
\multicolumn{1}{|c}{}                                    \\[-5dd]
\hline
\multicolumn{1}{r|}{\rule{15mm}{0dd}}                         &
\multicolumn{1}{|c|}{\rule{15mm}{0dd}}                        &
\multicolumn{1}{|c}{\rule{43mm}{0dd}}                         &
\multicolumn{1}{c|}{\rule{43mm}{0dd}}                         &
\multicolumn{1}{|c}{\rule{36mm}{0dd}}                  \\[-5dd]
\multicolumn{1}{r|}{\rule{15mm}{0dd}}                         &
\multicolumn{1}{|c|}{\rule{15mm}{0dd}}                        &
\multicolumn{1}{|c}{\rule{43mm}{0dd}}                         &
\multicolumn{1}{c|}{\rule{43mm}{0dd}}                         &
\multicolumn{1}{|c}{\rule{36mm}{0dd}}                  \\[-5dd]
\multicolumn{1}{c|}{ }                                          &
\multicolumn{1}{|c|}{ }                                         &
\multicolumn{1}{|r}{\begin{picture}(102,30)
\put(0,0){\line(1,0){40}}
\put(0,0){\line(0,1){24}}
\put(40,0){\line(0,1){24}}
\put(0,24){\line(1,0){40}}
\put(8,12){\line(1,0){92}}
\put(8,12){\circle*{3}}
\put(8,12){\circle{6}}
\put(31,12){\circle*{3}}
\put(54,12){\circle*{3}}
\put(77,12){\circle*{3}}
\put(100,12){\circle*{3}}
\end{picture}
\raisebox{8pt}{$l=0$}}                                      &
\multicolumn{1}{c|}{\begin{picture}(60,30)
\put(0,0){\line(1,0){50}}
\put(0,0){\line(0,1){30}}
\put(50,0){\line(0,1){30}}
\put(0,30){\line(1,0){50}}
\put(10,15){\line(1,0){23}}
\put(10,15){\circle*{3}}
\put(10,15){\circle{6}}
\put(33,15){\line(2,1){23}}
\put(33,15){\line(1,0){24}}
\put(57.8,15){\circle*{3}}
\put(33,15){\line(2,-1){23}}
\put(33,15){\circle*{3}}
\put(58,27){\circle*{3}}
\put(58,3){\circle*{3}}
\end{picture}
\raisebox{8pt}{$l=2$}}                                         &
\multicolumn{1}{|r}{}                \\
\multicolumn{1}{r|}{\rule{15mm}{0dd}}                         &
\multicolumn{1}{|c|}{\rule{15mm}{0dd}}                        &
\multicolumn{1}{|c}{\rule{43mm}{0dd}}                         &
\multicolumn{1}{c|}{\rule{43mm}{0dd}}                         &
\multicolumn{1}{|c}{\rule{36mm}{0dd}}                  \\[-5dd]
\multicolumn{1}{c|}{ }                                          &
\multicolumn{1}{|c|}{ }                                         &
\multicolumn{1}{|l}{\rule{8dd}{0dd}${\frak M}^{c}=3$}         &
\multicolumn{1}{l|}{\rule{26dd}{0dd}${\frak M}^{c}=5$}         &
\multicolumn{1}{|r}{\rule{8dd}{0dd}\fbox{${\bf C}_{v_{F}=4}$}}
\\ \multicolumn{1}{r|}{\rule{15mm}{0dd}}
& \multicolumn{1}{|c|}{\rule{15mm}{0dd}}                        &
\multicolumn{1}{|c}{\rule{43mm}{0dd}}                         &
\multicolumn{1}{c|}{\rule{43mm}{0dd}}                         &
\multicolumn{1}{|c}{\rule{36mm}{0dd}}                  \\[-5dd]
\multicolumn{1}{c|}{ }                                          &
\multicolumn{1}{|c|}{ }                                         &
\multicolumn{1}{|l}{\begin{picture}(84,30)
\put(0,0){\line(1,0){50}}
\put(0,0){\line(0,1){30}}
\put(50,0){\line(0,1){30}}
\put(0,30){\line(1,0){50}}
\put(10,15){\line(1,0){46}}
\put(10,15){\circle*{3}}
\put(10,15){\circle{6}}
\put(33,15){\circle*{3}}
\put(56,15){\circle*{3}}
\put(56,15){\line(2,1){23}}
\put(56,15){\line(2,-1){23}}
\put(56,15){\circle*{3}}
\put(81,27){\circle*{3}}
\put(81,3){\circle*{3}}
\end{picture}
\raisebox{12pt}{
$l=1$}} &
\multicolumn{1}{l|}{\rule{6mm}{0dd}\begin{picture}(62,30)
\put(0,0){\line(1,0){50}}
\put(0,0){\line(0,1){30}}
\put(50,0){\line(0,1){30}}
\put(0,30){\line(1,0){50}}
\put(10,15){\circle*{3}}
\put(10,15){\circle{6}}
\put(13,15){\line(3,1){23}}
\put(13,15){\line(3,-1){23}}
\put(38,23){\circle*{3}}
\put(38,7){\circle*{3}}
\put(38,23){\line(1,0){23}}
\put(38,7){\line(1,0){23}}
\put(61,23){\circle*{3}}
\put(61,7){\circle*{3}}
\end{picture}
\raisebox{12pt}{$l=2$}}                     &
\multicolumn{1}{|l}{\rule{3mm}{0dd}\begin{picture}(50,40)
\put(0,0){\line(1,0){50}}
\put(0,0){\line(0,1){40}}
\put(50,0){\line(0,1){40}}
\put(0,40){\line(1,0){50}}
\put(10,20){\circle*{3}}
\put(10,20){\circle{6}}
\put(10,20){\line(4,1){23}}
\put(10,20){\line(3,2){23}}
\put(10,20){\line(4,-1){23}}
\put(10,20){\line(3,-2){23}}
\put(33,26){\circle*{3}}
\put(33,35){\circle*{3}}
\put(33,15){\circle*{3}}
\put(33,6){\circle*{3}}
\end{picture}
 }                \\
\multicolumn{1}{r|}{\rule{15mm}{0dd}}                         &
\multicolumn{1}{|c|}{\rule{15mm}{0dd}}                        &
\multicolumn{1}{|c}{\rule{43mm}{0dd}}                         &
\multicolumn{1}{c|}{\rule{43mm}{0dd}}                         &
\multicolumn{1}{|c}{\rule{36mm}{0dd}}                  \\[-5dd]
\multicolumn{1}{c|}{ }                                          &
\multicolumn{1}{|c|}{ }                                         &
\multicolumn{1}{|l}{\rule{8dd}{0dd}${\frak M}^{c}=3$}         &
\multicolumn{1}{l|}{\rule{28dd}{0dd}${\frak M}^{c}=6$}         &
\multicolumn{1}{|l}{\rule{8dd}{0dd}${\frak M}^{c}_{\rm
max}=8$}                                                       \\
\multicolumn{1}{r|}{\rule{15mm}{0dd}}                         &
\multicolumn{1}{|c|}{\rule{15mm}{0dd}}                        &
\multicolumn{1}{|c}{\rule{43mm}{0dd}}                         &
\multicolumn{1}{c|}{\rule{43mm}{0dd}}                         &
\multicolumn{1}{|c}{\rule{36mm}{0dd}}                  \\[-5dd]
\multicolumn{1}{c|}{5}                                          &
\multicolumn{1}{|c|}{9}                                         &
\multicolumn{1}{|l}{\begin{picture}(62,30)
\put(0,0){\line(1,0){50}}
\put(0,0){\line(0,1){30}}
\put(50,0){\line(0,1){30}}
\put(0,30){\line(1,0){50}}
\put(10,15){\line(1,0){23}}
\put(10,15){\circle*{3}}
\put(10,15){\circle{6}}
\put(33,15){\line(2,1){23}}
\put(33,15){\line(2,-1){23}}
\put(33,15){\circle*{3}}
\put(58,27){\circle*{3}}
\put(58,3){\circle*{3}}
\put(58,27){\line(1,0){23}}
\put(81,27){\circle*{3}}
\end{picture}
$l=2$\raisebox{24pt}{$l=1$}}                                 &
\multicolumn{1}{c|}{\begin{picture}(62,40)
\put(0,0){\line(1,0){50}}
\put(0,0){\line(0,1){40}}
\put(50,0){\line(0,1){40}}
\put(0,40){\line(1,0){50}}
\put(10,20){\circle*{3}}
\put(10,20){\circle{6}}
\put(10,20){\line(3,1){23}}
\put(10,20){\line(3,-1){23}}
\put(35,28){\circle*{3}}
\put(35,11){\circle*{3}}
\put(35,28){\line(3,1){23}}
\put(35,28){\line(3,-1){23}}
\put(60,36){\circle*{3}}
\put(60,20){\circle*{3}}
\end{picture}
\raisebox{24pt}{$l=2$}}                                   &
\multicolumn{1}{|r}{ }                                       \\
\multicolumn{1}{r|}{\rule{15mm}{0dd}}                       &
\multicolumn{1}{|c|}{\rule{15mm}{0dd}}                        &
\multicolumn{1}{|c}{\rule{43mm}{0dd}}                         &
\multicolumn{1}{c|}{\rule{43mm}{0dd}}                         &
\multicolumn{1}{|c}{\rule{36mm}{0dd}}                  \\[-5dd]
\multicolumn{1}{c|}{ }                                          &
\multicolumn{1}{|c|}{ }                                         &
\multicolumn{1}{|r}{\rule{20dd}{0dd}}                    &
\multicolumn{1}{r|}{     }                                      &
\multicolumn{1}{|r}{                            }        \\[-5dd]
\multicolumn{1}{c|}{ }                                          &
\multicolumn{1}{|c|}{ }                                         &
\multicolumn{1}{|l}{\rule{8dd}{0dd}${\frak M}^{c}=4$}         &
\multicolumn{1}{l|}{\rule{28dd}{0dd}${\frak M}^{c}=6$}         &
\multicolumn{1}{|l}{     }                                     \\
\multicolumn{1}{r|}{\rule{15mm}{0dd}}                         &
\multicolumn{1}{|c|}{\rule{15mm}{0dd}}                        &
\multicolumn{1}{|c}{\rule{43mm}{0dd}}                         &
\multicolumn{1}{c|}{\rule{43mm}{0dd}}                         &
\multicolumn{1}{|c}{\rule{36mm}{0dd}}                        \\
\multicolumn{1}{c|}{ }                                          &
\multicolumn{1}{|c|}{ }                                         &
\multicolumn{1}{|l}{\begin{picture}(82,30)
\put(0,0){\line(1,0){50}}
\put(0,0){\line(0,1){30}}
\put(50,0){\line(0,1){30}}
\put(0,30){\line(1,0){50}}
\put(10,15){\circle*{3}}
\put(10,15){\circle{6}}
\put(10,15){\line(3,1){23}}
\put(10,15){\line(3,-1){23}}
\put(35,23){\circle*{3}}
\put(35,6){\circle*{3}}
\put(35,23){\line(1,0){46}}
\put(58,23){\circle*{3}}
\put(81,23){\circle*{3}}
\end{picture}
\raisebox{20pt}{$l=1$}}                                        &
\multicolumn{1}{c|}{\begin{picture}(58,30)
\put(0,0){\line(1,0){50}}
\put(0,0){\line(0,1){30}}
\put(50,0){\line(0,1){30}}
\put(0,30){\line(1,0){50}}
\put(10,15){\circle*{3}}
\put(10,15){\circle{6}}
\put(10,15){\line(1,0){23}}
\put(33,15){\circle*{3}}
\put(10,15){\line(2,1){23}}
\put(10,15){\line(2,-1){23}}
\put(33,26){\circle*{3}}
\put(33,4){\circle*{3}}
\put(33,26){\line(1,0){23}}
\put(56,26){\circle*{3}}
\end{picture}
\raisebox{23pt}{$l=2$}} &
\multicolumn{1}{|r}{ }                                         \\
\multicolumn{1}{r|}{\rule{15mm}{0dd}}                         &
\multicolumn{1}{|c|}{\rule{15mm}{0dd}}                        &
\multicolumn{1}{|c}{\rule{43mm}{0dd}}                         &
\multicolumn{1}{c|}{\rule{43mm}{0dd}}                         &
\multicolumn{1}{|c}{\rule{36mm}{0dd}}                  \\[-5dd]
\multicolumn{1}{c|}{ }                                          &
\multicolumn{1}{|c|}{ }                                         &
\multicolumn{1}{|l}{\rule{8dd}{0dd}${\frak M}^{c}=5$}         &
\multicolumn{1}{l|}{\rule{28dd}{0dd}${\frak M}^{c}=7$}         &
\multicolumn{1}{|l}{     }                                     \\
\multicolumn{1}{r|}{\rule{15mm}{0dd}}                         &
\multicolumn{1}{|c|}{\rule{15mm}{0dd}}                        &
\multicolumn{1}{|c}{\rule{43mm}{0dd}}                         &
\multicolumn{1}{c|}{\rule{43mm}{0dd}}                         &
\multicolumn{1}{|c}{\rule{36mm}{0dd}}                  \\[-5dd]
\multicolumn{1}{c|}{ }                                          &
\multicolumn{1}{|c|}{ }                                         &
\multicolumn{1}{|l}{                                  }         &
\multicolumn{1}{r|}{
\fbox{$\displaystyle l \equiv
\mathop{l_{QN}=0, 1, 2\rule{4mm}{0dd}}^{
\displaystyle
n_{QN} = v - 2 = 3}$}}                                          &
\multicolumn{1}{|l}{}                                    \\[-5dd]
\multicolumn{1}{r|}{\rule{15mm}{0dd}}                         &
\multicolumn{1}{|c|}{\rule{15mm}{0dd}}                        &
\multicolumn{1}{|c}{\rule{43mm}{0dd}}                         &
\multicolumn{1}{c|}{\rule{43mm}{0dd}}                         &
\multicolumn{1}{|c}{\rule{36mm}{0dd}}                        \\
\end{tabular}
\end{center} }

\vskip 14dd

One must mention still the second part of a pointed interval,
namely a rest interval 6$\leqslant ${\it v}$\leqslant $8, which
includes the further analogous {\bf RvT}~--- compositions for
atomic and nuclear quantum ({\it n}$_{QN}$, {\it l}$_{QN}$)~---
systems up to a case {\it v}=8, {\it T}$_{v=8}$=115 (corresponding
to a magic number for protons in nuclear shells) and also at {\it
v}=6 and $v$=7 for two light {\it u}- and {\it d}-quarks from
(1,0)-subsequence (see next section 2.8, Table 5).

At last from the very beginning of a consideration of the
superposition of different types of interaction it is necessary
to lay emphasis again on a possibility of existence of every
interaction for any microobject. Owing to this orientation we can
formulate a comprehensive superposition problem particularly for a
concrete microobject at {\it v}=11 with {\it T}$_{v}$=1842 (see
below Table 3). Nevertheless, if we assume that in a general case
the full {\it T}$_{v}$-set of non-directed and directed
microobject {\bf RvT}\,s, at a given number of vertices {\it v}, is
responsible for an all types of microobject interaction  then
naturally to introduce a division of {\it T}$_{v}$-set on the
corresponding {\bf RvT}-subsets: every of these {\bf
RvT}-subsets describes a definite type of interaction. In other
words, different {\bf RvT}s or different subsets of {\bf RvT} from
a full {\it T}$_{v}$-set of microobject {\bf RvT}\,s represent the
concrete sorts of interaction ``charge''. An essential role in the
distribution and arrangement of various {\bf RvT}\,s among the
different interactions of specific microobject may be attributed to
the properties of corresponding {\bf RvT}\,s ``core'' within which
for every ``free''~--- terminal~--- vertex ({\it v}$^{c}_{F}$) a
distance from root vertex ({\it v}$_{R}$) is equal to 1 and none of
these vertices, adding in root vertex, can be ``seen'' in isolation
i.~e.  we have a peculiar ``invisible ``core'' vertices
confinement'' into {\bf RvT}\,s (analogously with a ``quark
confinement'' hypothesis).

First of all one must separate from a full {\it T}$_{v}$-set of
{\bf RvT}\,s only the pure ``core'' {\bf RvT}\,s with
${\frak M}^{c}_{max}$=2({\it v}--1) and compare them with some
suitable interactions. Most probably that on the basis of the pure
``core'' {\bf RvT}\,s, due to ``vertices confinement'', may be
described namely the short-ranged or ``contact'' types of
interaction, i.~e. {\it strong and weak (with ``breaking'')
interactions} which operate {\it at very short range inside the
corresponding microobject}. In other words, we have in this case an
entirely ``microworld situation'' in which must figure only
typical discrete microobjects with ``invisible ``core'' vertices
confinement''.

On the other hand, {\it gravitational and electromagnetic
interactions} are {\it long-ranged and generally macroscopic} (in
space---time the strength of like force falls off with the square of
the distance between the interacting micro-, meso- and macroobjects).
At that one takes place a transition from ``microworld situation'',
reflected by the pure ``core'' {\bf RvT}\,s, to ``mixed or pure
macroworld situation'' where at first one or several {\bf
RvT}-vertices, beyond a ``core'', get ``visible'' and therefore
one arises the ``deconfinement'' of still more number of {\bf
RvT}-vertices, determined a construction of corresponding stable
physical systems (see, for example, shell systems of atoms and
nuclei in Table 2), their decays and reactions providing with
the Ulam's hypothesis.

\pagebreak
\begin{center}

{\bf Table~3. Superposition of
different types of interaction for concrete microobject with a
``core''-source of ``charges'' at {\it v}=11, {\it T}$_{v}$=1842.}

\end{center}

\vskip 4dd

{\tabcolsep=3dd
\begin{center}
\newcommand{\Sloppy}{\emergencystretch 3em \tolerance 9999
}
\let\sloppy=\Sloppy
\begin{tabular}{p{18mm}|p{30mm}|p{30mm}|p{18mm}|p{57mm}}
\hline
\multicolumn{1}{c|}{}                                           &
\multicolumn{2}{|c|}{}                                          &
\multicolumn{1}{|c|}{}                                          &
\multicolumn{1}{|c}{}                                    \\[-5dd]
\multicolumn{1}{c|}{Types of}                                   &
\multicolumn{2}{|c|}{{\bf RvT} --- subsets}                     &
\multicolumn{1}{|c|}{Number}                                    &
\multicolumn{1}{|c}{Systems, decays and reactions}             \\
\multicolumn{1}{c|}{interac-}                                   &
\multicolumn{2}{|c|}{with ${\frak M}^{c}$ --- intervals}        &
\multicolumn{1}{|c|}{of {\bf RvT}\,s}                           &
\multicolumn{1}{|c}{with corresponding root vertex}            \\
\multicolumn{1}{c|}{tion    }                                   &
\multicolumn{2}{|c|}{ }                                         &
\multicolumn{1}{|c|}{in             }                           &
\multicolumn{1}{|c}{``fission'' and ``fusion''}                \\
\multicolumn{1}{c|}{        }                                   &
\multicolumn{2}{|c|}{ }                                         &
\multicolumn{1}{|c|}{subsets        }                           &
\multicolumn{1}{|c}{                          }                \\
\multicolumn{1}{c|}{}                                           &
\multicolumn{2}{|c|}{}                                          &
\multicolumn{1}{|c|}{}                                          &
\multicolumn{1}{|c}{}                                    \\[-5dd]
\hline
\multicolumn{1}{r|}{\rule{18mm}{0dd}}                         &
\multicolumn{1}{|c}{\rule{30mm}{0dd}}                         &
\multicolumn{1}{c|}{\rule{30mm}{0dd}}                         &
\multicolumn{1}{|c|}{\rule{18mm}{0dd}}                        &
\multicolumn{1}{|c}{\rule{57mm}{0dd}}                  \\
\multicolumn{1}{c|}{ }                                          &
\multicolumn{1}{|l}{\begin{picture}(52,32)
\put(0,0){\line(1,0){50}}
\put(0,0){\line(0,1){30}}
\put(50,0){\line(0,1){30}}
\put(0,30){\line(1,0){50}}
\put(10,15){\line(1,0){50}}
\put(64,15){\circle*{1}}
\put(68,15){\circle*{1}}
\put(10,15){\circle*{3}}
\put(10,15){\circle{6}}
\put(40,15){\circle*{3}}
\end{picture}
\hfill\raisebox{12pt}{\ldots}}                                  &
\multicolumn{1}{l|}{\begin{picture}(58,30)
\put(0,0){\line(1,0){50}}
\put(0,0){\line(0,1){30}}
\put(50,0){\line(0,1){30}}
\put(0,30){\line(1,0){50}}
\put(10,15){\line(1,0){23}}
\put(10,15){\circle*{3}}
\put(10,15){\circle{6}}
\put(33,15){\line(2,1){23}}
\put(33,15){\line(2,-1){23}}
\put(33,15){\circle*{3}}
\put(60,29){\circle*{1}}
\put(64,31){\circle*{1}}
\put(60,2){\circle*{1}}
\put(64,0){\circle*{1}}
\end{picture}
\hfill\raisebox{12pt}{\ldots}}                                  &
\multicolumn{1}{|c|}{ }                                         &
\multicolumn{1}{|c}{                          }                \\
\multicolumn{1}{r|}{\rule{18mm}{0dd}}                         &
\multicolumn{1}{|c}{\rule{30mm}{0dd}}                         &
\multicolumn{1}{c|}{\rule{30mm}{0dd}}                         &
\multicolumn{1}{|c|}{\rule{18mm}{0dd}}                        &
\multicolumn{1}{|c}{\rule{57mm}{0dd}}                  \\[-5dd]
\multicolumn{1}{c|}{}                                         &
\multicolumn{1}{|l}{\rule{3mm}{0dd}${\frak M}^{c}=3$}        &
\multicolumn{1}{l|}{\rule{3mm}{0dd}${\frak M}^{c}=4$}         &
\multicolumn{1}{|c|}{}                                         &
\multicolumn{1}{|c}{      }                                \\
\multicolumn{1}{r|}{\rule{18mm}{0dd}}                         &
\multicolumn{1}{|c}{\rule{30mm}{0dd}}                         &
\multicolumn{1}{c|}{\rule{30mm}{0dd}}                         &
\multicolumn{1}{|c|}{\rule{18mm}{0dd}}                        &
\multicolumn{1}{|c}{\rule{57mm}{0dd}}                  \\[-5dd]
\multicolumn{1}{c|}{ }                                 &
\multicolumn{1}{|l}{\begin{picture}(58,30)
\put(0,0){\line(1,0){50}}
\put(0,0){\line(0,1){30}}
\put(50,0){\line(0,1){30}}
\put(0,30){\line(1,0){50}}
\put(10,15){\line(1,0){23}}
\put(10,15){\circle*{3}}
\put(10,15){\circle{6}}
\put(33,15){\line(2,1){23}}
\put(33,15){\line(1,0){24}}
\put(61,15){\circle*{1}}
\put(65,15){\circle*{1}}
\put(33,15){\line(2,-1){23}}
\put(33,15){\circle*{3}}
\put(60,29){\circle*{1}}
\put(64,31){\circle*{1}}
\put(60,2){\circle*{1}}
\put(64,0){\circle*{1}}
\end{picture}
\hfill\raisebox{14pt}{\ldots}} &
\multicolumn{1}{l|}{
\begin{picture}(62,30)
\put(0,0){\line(1,0){50}}
\put(0,0){\line(0,1){30}}
\put(50,0){\line(0,1){30}}
\put(0,30){\line(1,0){50}}
\put(10,15){\circle*{3}}
\put(10,15){\circle{6}}
\put(10,15){\line(3,1){23}}
\put(10,15){\line(3,-1){23}}
\put(35,23){\circle*{3}}
\put(35,6){\circle*{3}}
\put(35,23){\line(1,0){26}}
\put(64,23){\circle*{1}}
\put(68,23){\circle*{1}}
\end{picture}
\hfill\raisebox{14pt}{\ldots}}                     &
\multicolumn{1}{|c|}{ }                                         &
\multicolumn{1}{|c}{ }            \\
\multicolumn{1}{r|}{\rule{18mm}{0dd}}                         &
\multicolumn{1}{|c}{\rule{30mm}{0dd}}                         &
\multicolumn{1}{c|}{\rule{30mm}{0dd}}                         &
\multicolumn{1}{|c|}{\rule{18mm}{0dd}}                        &
\multicolumn{1}{|c}{\rule{57mm}{0dd}}                  \\[-5dd]
\multicolumn{1}{c|}{ }                             &
\multicolumn{1}{|l}{\rule{3mm}{0dd}${\frak M}^{c}=5$}           &
\multicolumn{1}{l|}{\rule{3mm}{0dd}${\frak M}^{c}=5$}           &
\multicolumn{1}{|c|}{1840}                                      &
\multicolumn{1}{|c}{      }                                \\
\multicolumn{1}{r|}{\rule{18mm}{0dd}}                         &
\multicolumn{1}{|c}{\rule{30mm}{0dd}}                         &
\multicolumn{1}{c|}{\rule{30mm}{0dd}}                         &
\multicolumn{1}{|c|}{\rule{18mm}{0dd}}                        &
\multicolumn{1}{|c}{\rule{57mm}{0dd}}                  \\[-5dd]
\multicolumn{1}{c|}{   }                                 &
\multicolumn{1}{|l}{
\begin{picture}(58,30)
\put(0,0){\line(1,0){50}}
\put(0,0){\line(0,1){30}}
\put(50,0){\line(0,1){30}}
\put(0,30){\line(1,0){50}}
\put(10,15){\line(1,0){23}}
\put(10,15){\circle*{3}}
\put(10,15){\circle{6}}
\put(33,15){\line(2,1){23}}
\put(33,15){\line(5,1){23}}
\put(33,15){\line(5,-1){23}}
\put(33,15){\line(2,-1){23}}
\put(33,15){\circle*{3}}
\put(61,21){\circle*{1}}
\put(65,22){\circle*{1}}
\put(61,9){\circle*{1}}
\put(65,8){\circle*{1}}
\put(60,29){\circle*{1}}
\put(64,31){\circle*{1}}
\put(60,2){\circle*{1}}
\put(64,0){\circle*{1}}
\end{picture}
\hfill\raisebox{14pt}{\ldots}} &
\multicolumn{1}{l|}{
\begin{picture}(63,30)
\put(0,0){\line(1,0){50}}
\put(0,0){\line(0,1){30}}
\put(50,0){\line(0,1){30}}
\put(0,30){\line(1,0){50}}
\put(10,15){\circle*{3}}
\put(10,15){\circle{6}}
\put(10,15){\line(3,1){23}}
\put(10,15){\line(3,-1){23}}
\put(35,23){\circle*{3}}
\put(35,6){\circle*{3}}
\put(35,23){\line(1,0){26}}
\put(35,6){\line(1,0){26}}
\put(64,23){\circle*{1}}
\put(68,23){\circle*{1}}
\put(64,6){\circle*{1}}
\put(68,6){\circle*{1}}
\end{picture}
\hfill\raisebox{14pt}{\ldots}}                     &
\multicolumn{1}{|c|}{ }                                         &
\multicolumn{1}{|c}{ }                            \\
\multicolumn{1}{r|}{\rule{18mm}{0dd}}                         &
\multicolumn{1}{|c}{\rule{30mm}{0dd}}                         &
\multicolumn{1}{c|}{\rule{30mm}{0dd}}                         &
\multicolumn{1}{|c|}{\rule{18mm}{0dd}}                        &
\multicolumn{1}{|c}{\rule{57mm}{0dd}}                  \\[-5dd]
\multicolumn{1}{c|}{ }                               &
\multicolumn{1}{|l}{${\rule{3mm}{0dd}\frak M}^{c}=6$}           &
\multicolumn{1}{l|}{${\rule{3mm}{0dd}\frak M}^{c}=6$}           &
\multicolumn{1}{|c|}{    }                                      &
\multicolumn{1}{|c}{ }         \\
\multicolumn{1}{r|}{\rule{18mm}{0dd}}                         &
\multicolumn{1}{|c}{\rule{30mm}{0dd}}                         &
\multicolumn{1}{c|}{\rule{30mm}{0dd}}                         &
\multicolumn{1}{|c|}{\rule{18mm}{0dd}}                        &
\multicolumn{1}{|c}{\rule{57mm}{0dd}}                  \\[-5dd]
\multicolumn{1}{c|}{ }                                           &
\multicolumn{1}{|l}{\begin{picture}(62,40)
\put(0,0){\line(1,0){50}}
\put(0,0){\line(0,1){40}}
\put(50,0){\line(0,1){40}}
\put(0,40){\line(1,0){50}}
\put(10,20){\circle*{3}}
\put(10,20){\circle{6}}
\put(10,20){\line(3,1){23}}
\put(10,20){\line(3,-1){23}}
\put(35,28){\circle*{3}}
\put(35,11){\circle*{3}}
\put(35,28){\line(3,1){23}}
\put(35,28){\line(3,-1){23}}
\put(61,19){\circle*{1}}
\put(65,18){\circle*{1}}
\put(61,37){\circle*{1}}
\put(65,39){\circle*{1}}
\end{picture}
\hfill\ldots,}                    &
\multicolumn{1}{l|}{etc.                  }                     &
\multicolumn{1}{|c|}{ }                                         &
\multicolumn{1}{|c}{              }                            \\
\multicolumn{1}{r|}{\rule{18mm}{0dd}}                         &
\multicolumn{1}{|c}{\rule{30mm}{0dd}}                         &
\multicolumn{1}{c|}{\rule{30mm}{0dd}}                         &
\multicolumn{1}{|c|}{\rule{18mm}{0dd}}                        &
\multicolumn{1}{|c}{\rule{57mm}{0dd}}                  \\[-5dd]
\multicolumn{1}{c|}{         }                                  &
\multicolumn{1}{|l}{\rule{3mm}{0dd}${\frak M}^{c}=6$}           &
\multicolumn{1}{c|}{                 }                          &
\multicolumn{1}{|c|}{    }                                      &
\multicolumn{1}{|c}{                                 }         \\
\multicolumn{1}{r|}{\rule{18mm}{0dd}}                         &
\multicolumn{1}{|c}{\rule{30mm}{0dd}}                         &
\multicolumn{1}{c|}{\rule{30mm}{0dd}}                         &
\multicolumn{1}{|c|}{\rule{18mm}{0dd}}                        &
\multicolumn{1}{|c}{\rule{57mm}{0dd}}                  \\[-5dd]
\multicolumn{1}{c|}{         }                                  &
\multicolumn{1}{|c}{\fbox{$3 \leqslant {\frak M}^{c}
\leqslant 18$},}                                                &
\multicolumn{1}{c|}{$1 \leqslant v^{c}_{F} \leqslant 7$}        &
\multicolumn{1}{|c|}{    }                                      &
\multicolumn{1}{|c}{                        }
\\[-325dd] 
\multicolumn{1}{c|}{Gravita-}                                 &
\multicolumn{2}{|c|}{}                                          &
\multicolumn{1}{|c|}{}                                          &
\multicolumn{1}{|c}{For stable}                     \\[5dd]
\multicolumn{1}{c|}{tional}                                 &
\multicolumn{2}{|c|}{}                                          &
\multicolumn{1}{|c|}{}                                          &
\multicolumn{1}{|c}{$(n_{QN}, l_{QN})$ --- systems}            \\ [5dd]
\multicolumn{1}{c|}{and electro-}                             &
\multicolumn{2}{|c|}{}                                          &
\multicolumn{1}{|c|}{}                                          &
\multicolumn{1}{|c}{with      }                                \\ [5dd]
\multicolumn{1}{c|}{magnetic  }                                 &
\multicolumn{2}{|c|}{}                                          &
\multicolumn{1}{|c|}{}                                          &
\multicolumn{1}{|c}{$n_{QN}=v-2=9$}                            \\ [5dd]
\multicolumn{1}{c|}{({\it long-}}                               &
\multicolumn{2}{|c|}{}                                          &
\multicolumn{1}{|c|}{}                                          &
\multicolumn{1}{|c}{$0 \leqslant l_{QN} \leqslant  8$}         \\ [5dd]
\multicolumn{1}{c|}{{\it ranged)}}              &
\multicolumn{2}{|c|}{}                                          &
\multicolumn{1}{|c|}{}                                          &
\multicolumn{1}{|c}{}                                    \\
\end{tabular}
\end{center} }
\newpage
\vskip 14dd

\noindent
{\tabcolsep=3dd
\begin{center}
\newcommand{\Sloppy}{\emergencystretch 3em \tolerance 9999
}
\let\sloppy=\Sloppy
\begin{tabular}{p{10mm}|p{20mm}|p{18mm}|p{10mm}|p{54mm}}
\hline
\multicolumn{1}{c|}{}                                           &
\multicolumn{2}{|c|}{}                                          &
\multicolumn{1}{|c|}{}                                          &
\multicolumn{1}{|c}{}                                    \\[-5dd]
\multicolumn{1}{c|}{Types of}                                   &
\multicolumn{2}{|c|}{{\bf RvT} --- subsets}                     &
\multicolumn{1}{|c|}{Number}                                    &
\multicolumn{1}{|c}{Systems, decays and reactions}             \\
\multicolumn{1}{c|}{interac-}                                   &
\multicolumn{2}{|c|}{with ${\frak M}^{c}$ --- intervals}        &
\multicolumn{1}{|c|}{of {\bf RvT}\,s}                           &
\multicolumn{1}{|c}{with corresponding root vertex}            \\
\multicolumn{1}{c|}{tion    }                                   &
\multicolumn{2}{|c|}{ }                                         &
\multicolumn{1}{|c|}{in             }                           &
\multicolumn{1}{|c}{``fission'' and ``fusion''}                \\
\multicolumn{1}{c|}{        }                                   &
\multicolumn{2}{|c|}{ }                                         &
\multicolumn{1}{|c|}{subsets        }                           &
\multicolumn{1}{|c}{                          }                \\
\multicolumn{1}{c|}{}                                           &
\multicolumn{2}{|c|}{}                                          &
\multicolumn{1}{|c|}{}                                          &
\multicolumn{1}{|c}{}                                    \\[-5dd]
\hline
\multicolumn{1}{r|}{\rule{10mm}{0dd}}                         &
\multicolumn{1}{|c}{\rule{20mm}{0dd}}                         &
\multicolumn{1}{c|}{\rule{18mm}{0dd}}                         &
\multicolumn{1}{|c|}{\rule{10mm}{0dd}}                        &
\multicolumn{1}{|c}{\rule{54mm}{0dd}}                        \\
\multicolumn{1}{c|}{         }                                  &
\multicolumn{1}{|l}{\begin{picture}(61,40)
\put(0,0){\line(1,0){50}}
\put(0,0){\line(0,1){40}}
\put(50,0){\line(0,1){40}}
\put(0,40){\line(1,0){50}}
\put(10,20){\circle*{3}}
\put(10,20){\circle{6}}
\put(10,20){\line(1,0){26}}
\put(10,20){\line(6,1){26}}
\put(10,20){\line(3,1){26}}
\put(10,20){\line(2,1){26}}
\put(10,20){\line(3,2){26}}
\put(10,20){\line(6,-1){26}}
\put(10,20){\line(3,-1){26}}
\put(10,20){\line(2,-1){26}}
\put(10,20){\line(3,-2){26}}
\put(36,20){\circle*{3}}
\put(36,24.5){\circle*{3}}
\put(36,28.5){\circle*{3}}
\put(36,32.5){\circle*{3}}
\put(36,36.5){\circle*{3}}
\put(36,16){\circle*{3}}
\put(36,11.5){\circle*{3}}
\put(36,7.5){\circle*{3}}
\put(36,3.5){\circle*{3}}
\put(36,3.5){\line(1,0){26}}
\put(62,3.5){\circle*{3}}
\end{picture}
\raisebox{-12pt}{\hspace{-50pt}$(n^{\circ})$}
\hfill\raisebox{1pt}{$\rightarrow$}}                &
\multicolumn{1}{l|}{\rule{3mm}{0dd}\begin{picture}(40,30)
\put(0,0){\line(0,1){30}}
\put(0,15){\circle*{3}}
\put(0,15){\circle{6}}
\put(0,30){\line(1,0){40}}
\put(12,30){\vector(1,0){10}}
\put(0,0){\line(1,0){40}}
\put(27,0){\vector(-1,0){10}}
\put(40,0){\line(0,1){30}}
\put(40,15){\circle*{3}}
\end{picture}
\raisebox{-12pt}{\hspace{-30pt}$(e^{+})$}
\raisebox{36pt}{\hspace{-25pt}$(e^{-})$} }                  &
\multicolumn{1}{|c|}{ }                                     &
\multicolumn{1}{|l}{\begin{picture}(61,40)
\put(0,0){\line(1,0){50}}
\put(0,0){\line(0,1){40}}
\put(50,0){\line(0,1){40}}
\put(0,40){\line(1,0){50}}
\put(10,20){\circle*{3}}
\put(10,20){\circle{6}}
\put(10,20){\line(1,0){26}}
\put(10,20){\line(6,1){26}}
\put(10,20){\line(3,1){26}}
\put(10,20){\line(2,1){26}}
\put(10,20){\line(3,2){26}}
\put(10,20){\line(6,-1){26}}
\put(10,20){\line(3,-1){26}}
\put(10,20){\line(2,-1){26}}
\put(10,20){\line(3,-2){26}}
\put(36,20){\circle*{3}}
\put(36,24.5){\circle*{3}}
\put(36,28.5){\circle*{3}}
\put(36,32.5){\circle*{3}}
\put(36,36.5){\circle*{3}}
\put(36,16){\circle*{3}}
\put(36,11.5){\circle*{3}}
\put(36,7.5){\circle*{3}}
\put(36,3.5){\circle*{3}}
\put(62,3.5){\vector(-1,0){26}}
\put(62,3.5){\circle*{3}}
\end{picture}
\rlap{\raisebox{-11pt}{\hspace{-55dd}$(p^{+})$\hspace{18dd}$(e^{+})$}}
\rlap{\raisebox{46pt}{\hspace{-50dd}For neutron's decay $n^{\circ}
\to p^{+} + e^{-} + \bar{\nu}_{e}$}}
\raisebox{14pt}{+}\hspace{5dd}\begin{picture}(52,32)
\put(0,0){\line(1,0){50}}
\put(0,0){\line(0,1){30}}
\put(50,0){\line(0,1){30}}
\put(0,30){\line(1,0){50}}
\put(10,15){\vector(1,0){30}}
\put(10,15){\circle*{3}}
\put(10,15){\circle{6}}
\put(40,15){\circle*{3}}
\end{picture}
\rlap{\raisebox{-11pt}{\hspace{-40dd}$(e^{-})$}}
\raisebox{14pt}{+}\hspace{5dd}\begin{picture}(50,40)
\put(0,0){\line(1,0){50}}
\put(0,0){\line(0,1){40}}
\put(50,0){\line(0,1){40}}
\put(0,40){\line(1,0){50}}
\put(10,20){\circle*{3}}
\put(10,20){\circle{6}}
\put(10,20){\vector(2,1){30}}
\put(40,5){\vector(-2,1){30}}
\put(40,35){\circle*{3}}
\put(40,5){\circle*{3}}
\end{picture}
\rlap{\raisebox{-11pt}{\hspace{-30dd}$(\bar{\nu}_{e})$}}
}         \\
\multicolumn{1}{c|}{Weak     }                                  &
\multicolumn{1}{|l}{   }                                      &
\multicolumn{1}{r|}{}                                         &
\multicolumn{1}{|c|}{1   }                                    &
\multicolumn{1}{|l}{
where $\displaystyle \bar{\nu}_{e}$ is identical to
$\displaystyle {\bf C}_{v_{F}=2}$ }                           \\
\multicolumn{1}{c|}{(\it short-}                              &
\multicolumn{1}{|l}{   }                                      &
\multicolumn{1}{r|}{}                                         &
\multicolumn{1}{|c|}{   }                                     &
\multicolumn{1}{|l}{with a pair $e^{-} - e^{+}$ or to a binding}  \\
\multicolumn{1}{c|}{\it ranged)}                 &
\multicolumn{2}{|c|}{\fbox{${\frak M}^{c}=19$}, \rule{4mm}{0mm}
$v^{c}_{F}=8$\rule{18mm}{0mm}   }
&
\multicolumn{1}{|c|}{}                                       &
\multicolumn{1}{|l}{energy of neutron $n^{\circ}$ (variable)}
\\
\multicolumn{1}{c|}{         }                                  &
\multicolumn{2}{c|}{\rule{20dd}{0dd}{\it OR}}                 &
\multicolumn{1}{|c|}{    }                                      &
\multicolumn{1}{|l}{ }  \\
\multicolumn{1}{c|}{         }                                  &
\multicolumn{2}{c|}{\rule{65dd}{0dd}$\to$ \
\begin{picture}(40,30)
\put(0,0){\line(0,1){30}}
\put(0,15){\circle*{3}}
\put(0,15){\circle{6}}
\put(0,30){\line(1,0){40}}
\put(12,30){\vector(1,0){10}}
\put(0,0){\line(1,0){40}}
\put(27,0){\vector(-1,0){10}}
\put(40,0){\line(0,1){30}}
\put(40,15){\circle*{3}}
\end{picture}
\raisebox{-12pt}{\hspace{-28pt}$(W^{+})$}
\raisebox{36pt}{\hspace{-29pt}$(W^{-})$} }
&
\multicolumn{1}{|c|}{    }                                      &
\raisebox{36pt}{ }       \\
\multicolumn{1}{c|}{         }                                  &
\multicolumn{2}{c|}{\rule{20dd}{0dd}etc.}                     &
\multicolumn{1}{|c|}{    }                                      &
                       \\
\multicolumn{1}{r|}{\rule{10mm}{0dd}}                         &
\multicolumn{1}{|c}{\rule{20mm}{0dd}}                         &
\multicolumn{1}{c|}{\rule{18mm}{0dd}}                         &
\multicolumn{1}{|c|}{\rule{10mm}{0dd}}                        &
\multicolumn{1}{|c}{\rule{54mm}{0dd}}                  \\[-5dd]
\hline
\multicolumn{1}{r|}{\rule{10mm}{0dd}}                         &
\multicolumn{1}{|c}{\rule{20mm}{0dd}}                         &
\multicolumn{1}{c|}{\rule{18mm}{0dd}}                         &
\multicolumn{1}{|c|}{\rule{10mm}{0dd}}                        &
\multicolumn{1}{|c}{\rule{54mm}{0dd}}                  \\[-5dd]
\multicolumn{1}{c|}{         }                                  &
\multicolumn{1}{|l}{\begin{picture}(50,42)
\put(0,0){\line(1,0){50}}
\put(0,0){\line(0,1){42}}
\put(50,0){\line(0,1){42}}
\put(0,42){\line(1,0){50}}
\put(10,19){\circle*{3}}
\put(10,19){\circle{6}}
\put(10,19){\line(1,0){26}}
\put(10,19){\line(6,1){26}}
\put(10,19){\line(3,1){26}}
\put(10,19){\line(2,1){26}}
\put(10,19){\line(3,2){26}}
\put(10,19){\line(5,4){26}}
\put(10,19){\line(6,-1){26}}
\put(10,19){\line(3,-1){26}}
\put(10,19){\line(2,-1){26}}
\put(10,19){\line(3,-2){26}}
\put(36,19){\circle*{3}}
\put(36,23){\circle*{3}}
\put(36,27.5){\circle*{3}}
\put(36,31.5){\circle*{3}}
\put(36,35.5){\circle*{3}}
\put(36,39.1){\circle*{3}}
\put(36,15){\circle*{3}}
\put(36,10.5){\circle*{3}}
\put(36,6.5){\circle*{3}}
\put(36,2.5){\circle*{3}}
\end{picture}
}                                                               &
\multicolumn{1}{c|}{\raisebox{22pt}{\fbox{{\bf C}$_{v_{F}=10}$}}}
                                                             &
\multicolumn{1}{|c|}{1}                                      &
\multicolumn{1}{|l}{\raisebox{0pt}{For schemes of various decays
and reactions}} \\
\multicolumn{1}{c|}{Strong   }                                  &
\multicolumn{1}{|l}{}                   &
\multicolumn{1}{r|}{}                &
\multicolumn{1}{|c|}{ }                                      &
\multicolumn{1}{|l}{\rule{3mm}{0dd}${\bf C}_{v_{F}=10}
\rightleftarrows 2 {\bf C}_{v_{F}=5},$}
\\
\multicolumn{1}{c|}{({\it short-}}
& \multicolumn{1}{|l}{                        }                   &
\multicolumn{1}{r|}{                           }                &
\multicolumn{1}{|c|}{    }                                      &
\multicolumn{1}{|l}{\rule{16mm}{0dd}$\rightleftarrows
{\bf C}_{v_{F}=5}+{\bf C}_{v_{F}=3}+{\bf C}_{v_{F}=2},$}       \\
\multicolumn{1}{c|}{{\it ranged})}                              &
\multicolumn{2}{|l}{\fbox{$\displaystyle{\frak M}^{c} = 2 (v -
1) = 20$}\,,}                                                 &
\multicolumn{1}{|c|}{    }                                      &
\multicolumn{1}{|l}{\rule{16mm}{0dd}$\rightleftarrows
2{\bf C}_{v_{F}=4}+{\bf C}_{v_{F}=2},$}                        \\
\multicolumn{1}{r|}{\rule{10mm}{0dd}}                         &
\multicolumn{1}{|c}{\rule{20mm}{0dd}}                         &
\multicolumn{1}{c|}{ }                         &
\multicolumn{1}{|c|}{\rule{10mm}{0dd}}                        &
\multicolumn{1}{|c}{\rule{54mm}{0dd}}                  \\[-5dd]
\multicolumn{1}{c|}{}                                            &
\multicolumn{1}{|l}{\rule{3mm}{0dd}$v^{c}_{F} = v - 1 =10$}     &
\multicolumn{1}{c|}{ }                                      &
\multicolumn{1}{|l|}{ }                                      &
etc. adequate to root vertex-corresponding superposition      \\
\end{tabular}
\end{center} }

\vskip 14dd

It is interesting that for $\overline{\nu}_{e}$-represented
{\bf C}$_{v_{F}=2}$ an incidence matrix $\displaystyle
\mathop{\bf I}_{(2 \times 2)} (\epsilon) = \left(
\begin{array}{cc} 1 & 0
\\ 0 & -1 \end{array}  \right)$ corresponds to one of Pauli's
spin-matrices.

Thus, in the framework of many-``planes'' {\bf
RvT}---representation of every known interaction of microobjects
one occur some typical {\bf RvT}---compositions with different sets
of ``free''---terminal---vertices ({\it v}$_{F}$), so-called
``hedgehogs''. Characteristically, that the short-ranged
interactions, as against the long-ranged ones, can be described
by a single ``hedgehog''~--- {\bf RvT} i.~e. for this case there is
an one-``plane'' {\bf RvT}~--- representation which must reflect at
the same time a ``compact'' inner structure of microobject in
respect to a given interaction. On the contrary, in a case of the
long-ranged interactions there are many ``hedgehogs''~--- {\bf
RvT}s determined to a large degree ``transparent'' inner structure
of corresponding physical systems having also many-``planes''
{\bf RvT}~--- representation. Such ``transparent'' inner structure
can be considered as an adequate hierarchical scheme of levels or
``orbitals'' with the different sets of ``hedgehog''~--- {\bf RvT}
(for this aspect see also the Table 2 and the following section
2.7).

It is noteworthy that a superposition of four (or more) types of
interaction occurs here necessarily as a result of the compulsory
``stratification'' of a corresponding set of ``hedgehogs''~---
{\bf RvT}\,s. Only after the superposition of different types of
interaction one can regard wholly any microobject with
many-``planes'' ``hedgehog''~--- {\bf RvT} representation, that must
naturally solve the problem of the whole and the parts within the
framework of an immanent construction of many-``charge''
microobject.

However, in this connection it is significant to emphasize that a
central problem of contemporary unified~--- field theory, that
would relate the electromagnetic, gravitational, strong, and weak
interactions in one set of equations and no like solution has yet
been found (some progress has been made in the unification of the
electromagnetic and weak interactions), now transforms to the
completely new problem of decoding and interpretation of a
many-``charge'' discrete microobject as an indivisible set of
``stratification planes'' in the frame of a ``hedgehog''~--- {\bf
RvT} formalism.

\subsection*{2.7 Atomic events in the new graph kinematics
formalism}

As it was emphasized by Bohr and Einstein in due time (see
Heisenberg's statement in Ref.~[20]) a stability of atom means a
permanent reproduction (an immediate mechanism) of the same
stationary forms or steady states in the framework of which

---~according to Bohr's postulates, we have an electron moving
about in certain orbit of atom and occasionally making a jump from
one orbit to another; in agreement with Bohr---Heisenberg's opinion
(see Refs. [20, 21]) it is impossible to describe similar jump in
the old notions as a discrete process in space---time,

---~following to quantum mechanics, we have in atoms the so-called
``orbitals'' i.~e. the regions of space where it is not possible to
give a definite path for an electron but only a probability
distribution or an electric charge distribution (averaged over
time) around the nucleus; one strong Einstein's objection(see again
Refs. [20, 21]) consists in the opposition of this conclusion to
the real paths of electrons in the Wilson cloud chamber.

It is known that both very different and naturally interconnected
aspects of these phenomena, reflected the unexpected {\it
stability} (a reproduction of the same steady states) and the
evident {\it discontinuity} (an electron jump between orbits) of
atomic system dynamics, are presented mathematically in
contemporary quantum mechanics of atom which allows

---~to calculate the discrete energy levels of stationary states,

---~to take into account an inexplicable stability of the
``orbitals'' or the definite stationary states which are
distinguished one from another in respect of their shapes and each
could be reproduced formally again and again,

---~to estimate a probability of the sudden emission of photon as a
result of the discrete transition between stationary states.

Nevertheless, according to Einstein (see Ref. [20]) in spite of
these doubtless achievements, many problems, especially the problem
of a transition itself between stationary states (without regard
for the real path of electrons) and the problem of a sudden
emission (or ``production'') of discrete photon, originated as the
results  of an analysis of the purely discontinuous atomic events,
haven't any satisfactory explanation in quantum mechanics and
demand the more accurate description. Unfortunately, it is easy to
set up that de facto these old unresolved, but clearly formulated
principal problems of quantum mechanics were thrown off at the
stage of a branching of fundamental physical theory in a middle of
the twentieth century in favour of some alternative variants of
development that lie now into the mainstream of current research.

This dead end for similar comprehensive conceptual problems give
rise to the many inappropriate latent difficulties in some new
developing physical models including the widely known divergence
difficulties. Therefore it is very notable that similar
discontinuous events may be described immediately on the base of
graph formalism for discrete atomic systems beyond space---time.
Indeed, the unconditional introduction of a such discrete graph
technique provides for the various subnuclear microobjects some
{\it single sets} of the concrete {\bf RvT} s at a given {\it v}
with the aim of a determination of their masses, directly
proportional to $T_{v}$~--- numbers or their fractional
($r,\,s$)~--- ``fragments'' $T_{v}/2^{r}3^{s}$. While a structure
of the stable atomic and nuclear shell (many-level) systems can be
described certainly with the help of the {\it several sets} of {\bf
RvT} s corresponding already to the different number of vertices
$v$ and thereby to the different principal quantum numbers
$n_{QN}=v-2$. At that in every {\bf RvT} these $v$ vertices must
have a definite disposition which additionally may be presented in
some alternative configurations, producing the ``graph isomers''
with the same number of vertices ($v_{i}$) but the different values
of their powers ${\frak M}_{i}$
(identical to their degrees or valencies, according to (3) and
(5)).

Naturally the so-called ``valence-isomerism'' will be realized by
means of the partitions (see Ref. [12]) and their representative
trees i.~e. {\bf RvT} s. Factually, the realizability of a set of
integers (in spite of ``fractals'') as degrees of the vertices of a
specific {\bf RvT} with the detachment of a root vertex ($v_{R}$)
and a ``core'' can be illustrated par example on the next
partitions for ($v=5, \ n=4$)~--- {\bf RvT} from Table~2 (with the
completeness~--- number on the left):

\vskip 0dd
{\tabcolsep=6dd
\begin{center}
\newcommand{\Sloppy}{\emergencystretch 3em \tolerance 9999
}
\let\sloppy=\Sloppy
\begin{tabular}{cc}
8=(1+2)+2+2+1                                            &
8=(1+4)+1+1+1                                           \\
8=(1+2)+3+1+1                                            &
8=(2+2+2)+1+1                                           \\
8=(1+3)+2+1+1                                            &
8=(2+3+1)+1+1                                            \\
8=(2+2+1)+2+1                                            &
8=(3+2+1+1)+1                                           \\
\multicolumn{2}{c}{8=(4+1+1+1+1).}
\end{tabular}
\end{center} }
\vskip 0dd

\noindent
Here we apply on the base of (5) a known theorem about the
tree-realization of the partitions (see Refs. [12, 19]):
$$2n\equiv 2(v-1)=\sum\limits^{v}_{i=1}{\bf
deg}v_{i}\equiv\sum\limits^{v}_{i=1}{\frak M}(v_{i}). \eqno(7)$$

\noindent
The complex of {\bf RvT} s in Table~2 is responsible, in
particular, for a three-level shell system with the principal
quantum number $n_{QN}=v-2$, equal to 1, 2 and 3, and with the
corresponding possible values for the orbital quantum number
$l_{QN}$:0, 1 and 2.

Then an above-pointed ``valence-isomerism'' reflects a
many-``planes'' representation of the like hierarchical shell
systems as atoms and nuclei (several {\bf RvT}~--- sets) that is
simply an expression of their stability. Therefore just the
simultaneous or ``parallel'' presence of all ``planes'' or
stationary forms of any stable shell system in {\bf RvT}'s
description means its permanent reproduction by the way of non-open
``transferability'' between ``graph isomers'' with the same number
of vertices. Thus one obtains an adequate interpretation of the
{\bf RvT}~--- representation of steady states for every shell
system where may be taken into account additionally a possible
appearance of the degenerate levels. Still more it seems very
significant also that the indicated {\bf RvT}~--- representation is
congruous with the corresponding one in Table~1 which describes the
real path of free electron, particularly in the Wilson cloud
chamber (beyond atom).

\subsubsection*{2.7.1 Graph schemes for elementary processes}

Now the transitions or ``jumps'' between such steady states of the
shell systems become absolutely evident as occuring with an
exchange of lines in their {\bf RvT}s and corresponding oriented
multigraphs. They take place in the next concrete examples which
can be considered by the special ``graph schemes'' or ``graph
equations'' for various atomic and nuclear elementary processes.

The graph scheme for an emission of photon by atom with jump down
of electron contains the loss of one of the peripheral vertex
($v_{p}$) in pointed below suitable {\bf RvT} from a collection of
subtrees {\bf T}--{\bf v}$_{p}$, corresponding to Bondy's condition
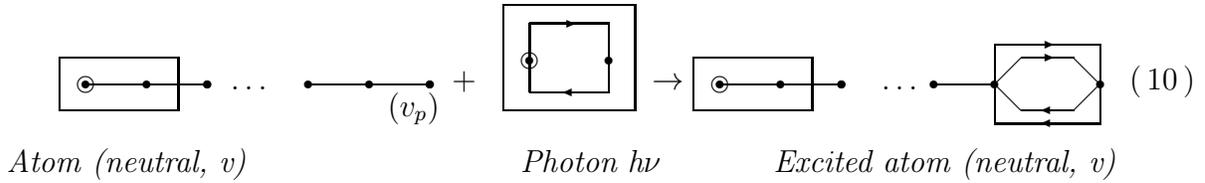
$$e_{T}(v_{p})=D_{T}\eqno(8)$$

\noindent
in agreement with (4a) and (4b):

\vskip 8dd
\rule{7mm}{0mm}
\begin{picture}(58,20)
\put(0,0){\line(1,0){45}}
\put(0,0){\line(0,1){20}}
\put(45,0){\line(0,1){20}}
\put(0,20){\line(1,0){45}}
\put(10,10){\line(1,0){46}}
\put(10,10){\circle*{3}}
\put(10,10){\circle{6}}
\put(33,10){\circle*{3}}
\put(56,10){\circle*{3}}
\put(56,10){\circle*{3}}
\end{picture}
\rule{1.2mm}{0mm}\raisebox{8pt}{\ldots}
\begin{picture}(58,20)
\put(10,10){\line(1,0){46}}
\put(40,-2){($v_p$)}
\put(10,10){\circle*{3}}
\put(33,10){\circle*{3}}
\put(56,10){\circle*{3}}
\put(56,10){\circle*{3}}
\end{picture}
\rule{1mm}{0mm}\raisebox{8pt}{$\rightarrow$}
\begin{picture}(58,20)
\put(0,0){\line(1,0){45}}
\put(0,0){\line(0,1){20}}
\put(45,0){\line(0,1){20}}
\put(0,20){\line(1,0){45}}
\put(10,10){\line(1,0){46}}
\put(10,10){\circle*{3}}
\put(10,10){\circle{6}}
\put(33,10){\circle*{3}}
\put(56,10){\circle*{3}}
\put(56,10){\circle*{3}}
\end{picture}
\rule{2mm}{0mm}\raisebox{8pt}{\ldots}
\rule{2mm}{0mm}
\begin{picture}(26,20)
\put(0,10){\line(1,0){23}}
\put(0,10){\circle*{3}}
\put(23,10){\circle*{3}}
\end{picture}
\rule{5mm}{0mm}\raisebox{8pt}{+}
\rule{1mm}{0mm}
\begin{picture}(50,40)
\put(0,0){\line(1,0){50}}
\put(0,0){\line(0,1){40}}
\put(50,0){\line(0,1){40}}
\put(0,40){\line(1,0){50}}
\put(10,7){\line(0,1){26}}
\put(10,19){\circle*{3}}
\put(10,19){\circle{6}}
\put(10,33){\line(1,0){30}}
\put(18,33){\vector(1,0){10}}
\put(10,7){\line(1,0){30}}
\put(32,7){\vector(-1,0){10}}
\put(40,7){\line(0,1){26}}
\put(40,19){\circle*{3}}
\end{picture}
\ \hfill\raisebox{8pt}{(\,9\,)}

\vskip 8dd

\noindent
{\it\rule{5mm}{0mm}Atom (neutral, v)\rule{30mm}{0mm}Atom (neutral,
v-1) \rule{25mm}{0mm} Photon h$\nu$}
\vskip 8dd

\noindent
In (9) neutral atom with $v-1$ vertices can be reconstructed by
means of an use of Ulam's hypothesis (according to Bondy) for {\bf
T}--{\bf v}$_{p}$ up to initial neutral atom with $v$ vertices
corresponding to {\bf T}. At that we have a production of photon
with an emergence of the root vertex and a photon ``core''
formation (in the h$\nu$~--- semicycle form).

It is easy to understand that an use of the directed {\bf RvT}s
for describing of concrete processes with the charged microobjects
instead of the non-directed ones for an estimation of microobject
masses may be carried out by the way of a treatment of the totally
or partially enumerative {\bf RvT}s i. e. {\bf RvT}s already with
marked vertices (in particular, with  $\displaystyle (v_{p})$~---
vertex in graph scheme (9)). Naturally, in any single case we have
a conservation of the total number of non-directed and cycle paar
of opposite directed lines for enumerative {\bf RvT}s in these
graph schemes; all over there are balanced signed digraphs.

The graph scheme for an absorption of photon by atom with jump up
of electron contains the result of an insertion of the photon's
h$\nu$~--- semicycle into the peripheral vertex $(v_{p})$
accompanied with a doubling~--- operation for the peripheral line:

\pagebreak
\rule{6mm}{0mm}
\begin{picture}(58,20)
\put(0,0){\line(1,0){45}}
\put(0,0){\line(0,1){20}}
\put(45,0){\line(0,1){20}}
\put(0,20){\line(1,0){45}}
\put(10,10){\line(1,0){46}}
\put(10,10){\circle*{3}}
\put(10,10){\circle{6}}
\put(33,10){\circle*{3}}
\put(56,10){\circle*{3}}
\put(56,10){\circle*{3}}
\end{picture}
\rule{1.2mm}{0mm}\raisebox{8pt}{\ldots}
\begin{picture}(58,20)
\put(10,10){\line(1,0){46}}
\put(40,-2){($v_p$)}
\put(10,10){\circle*{3}}
\put(33,10){\circle*{3}}
\put(56,10){\circle*{3}}
\put(56,10){\circle*{3}}
\end{picture}
\rule{1mm}{0mm}\raisebox{8pt}{+}
\rule{1mm}{0mm}
\begin{picture}(50,40)
\put(0,0){\line(1,0){50}}
\put(0,0){\line(0,1){40}}
\put(50,0){\line(0,1){40}}
\put(0,40){\line(1,0){50}}
\put(10,7){\line(0,1){26}}
\put(10,19){\circle*{3}}
\put(10,19){\circle{6}}
\put(10,33){\line(1,0){30}}
\put(18,33){\vector(1,0){10}}
\put(10,7){\line(1,0){30}}
\put(32,7){\vector(-1,0){10}}
\put(40,7){\line(0,1){26}}
\put(40,19){\circle*{3}}
\end{picture}
\rule{1mm}{0mm}\raisebox{8pt}{$\rightarrow$}
\raisebox{-0pt}{%
\begin{picture}(58,20)
\put(0,0){\line(1,0){45}}
\put(0,0){\line(0,1){20}}
\put(45,0){\line(0,1){20}}
\put(0,20){\line(1,0){45}}
\put(10,10){\line(1,0){46}}
\put(10,10){\circle*{3}}
\put(10,10){\circle{6}}
\put(33,10){\circle*{3}}
\put(56,10){\circle*{3}}
\put(56,10){\circle*{3}}
\end{picture}
}
\rule{2mm}{0mm}\raisebox{8pt}{\ldots}
\rule{7mm}{0mm}
\raisebox{-5pt}{%
\begin{picture}(40,30)
\put(0,0){\line(0,1){30}}
\put(0,15){\circle*{3}}
\put(0,30){\line(1,0){40}}
\put(0,0){\line(1,0){40}}
\put(40,0){\line(0,1){30}}
\put(40,15){\circle*{3}}
\put(0,15){\line(-1,0){23}}
\put(-23,15){\circle*{3}}
\put(13,25.1){\vector(1,0){10}}
\put(13,30){\vector(1,0){10}}
\put(0,15){\line(1,1){10}}
\put(0,15){\line(1,-1){10}}
\put(40,15){\line(-1,1){10}}
\put(40,15){\line(-1,-1){10}}
\put(10,25.1){\line(1,0){20}}
\put(10,5){\line(1,0){20}}
\put(27,5){\vector(-1,0){10}}
\put(27,0){\vector(-1,0){10}}
\end{picture}
}
\ \hfill\raisebox{8pt}{(\,10\,)}

\vskip8dd
\noindent
{\it\rule{5mm}{0mm} Atom (neutral, v)
\rule{35mm}{0mm}Photon h$\nu$
\rule{12mm}{0mm} Excited atom
(neutral, v)}
\vskip8dd

\noindent
In (10) we have on the right an excited neutral atom with $v$
vertices where the peripheral line is transformed to a neutral
signed semiwalk \begin{picture}(40,30)
\put(0,0){\line(0,1){30}}
\put(0,30){\line(1,0){40}}
\put(0,0){\line(1,0){40}}
\put(40,0){\line(0,1){30}}
\put(13,25.1){\vector(1,0){10}}
\put(13,30){\vector(1,0){10}}
\put(0,15){\line(1,1){10}}
\put(0,15){\line(1,-1){10}}
\put(40,15){\line(-1,1){10}}
\put(40,15){\line(-1,-1){10}}
\put(10,25.1){\line(1,0){20}}
\put(10,5){\line(1,0){20}}
\put(27,5){\vector(-1,0){10}}
\put(27,0){\vector(-1,0){10}}
\end{picture}
presented an excited state of atom.

At last we consider the graph scheme for photoelectric effect i. e.
for the liberation of electron from a substance atom exposed to
electromagnetic radiation:

\noindent
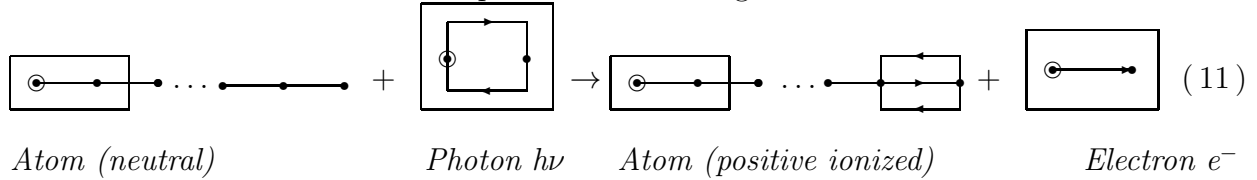
\begin{picture}(58,20)
\put(0,0){\line(1,0){45}}
\put(0,0){\line(0,1){20}}
\put(45,0){\line(0,1){20}}
\put(0,20){\line(1,0){45}}
\put(10,10){\line(1,0){46}}
\put(10,10){\circle*{3}}
\put(10,10){\circle{6}}
\put(33,10){\circle*{3}}
\put(56,10){\circle*{3}}
\put(56,10){\circle*{3}}
\end{picture}
\raisebox{8pt}{\ldots}
\raisebox{9pt}{%
\begin{picture}(48,3)
\put(0,0){\line(1,0){46}}
\put(0,0){\circle*{3}}
\put(23,0){\circle*{3}}
\put(46,0){\circle*{3}}
\end{picture}
}
\rule{1pt}{0mm}\raisebox{8pt}{+}
\rule{1mm}{0mm}
\begin{picture}(50,40)
\put(0,0){\line(1,0){50}}
\put(0,0){\line(0,1){40}}
\put(50,0){\line(0,1){40}}
\put(0,40){\line(1,0){50}}
\put(10,7){\line(0,1){26}}
\put(10,19){\circle*{3}}
\put(10,19){\circle{6}}
\put(10,33){\line(1,0){30}}
\put(18,33){\vector(1,0){10}}
\put(10,7){\line(1,0){30}}
\put(32,7){\vector(-1,0){10}}
\put(40,7){\line(0,1){26}}
\put(40,19){\circle*{3}}
\end{picture}
\rule{1mm}{0mm}\raisebox{8pt}{$\rightarrow$}
\begin{picture}(58,20)
\put(0,0){\line(1,0){45}}
\put(0,0){\line(0,1){20}}
\put(45,0){\line(0,1){20}}
\put(0,20){\line(1,0){45}}
\put(10,10){\line(1,0){46}}
\put(10,10){\circle*{3}}
\put(10,10){\circle{6}}
\put(33,10){\circle*{3}}
\put(56,10){\circle*{3}}
\put(56,10){\circle*{3}}
\end{picture}
\rule{1mm}{0mm}\raisebox{8pt}{\ldots}
\rule{5.5mm}{0mm}
\begin{picture}(30,20)
\put(0,0){\line(0,1){20}}
\put(0,20){\line(1,0){30}}
\put(0,0){\line(1,0){30}}
\put(30,0){\line(0,1){20}}
\put(0,10){\circle*{3}}
\put(0,10){\line(-1,0){20}}
\put(-20,10){\circle*{3}}
\put(30,10){\circle*{3}}
\put(0,10){\line(1,0){30}}
\put(22,0){\vector(-1,0){10}}
\put(7,10){\vector(1,0){10}}
\put(22,20){\vector(-1,0){10}}
\end{picture}
\rule{1mm}{0mm}\raisebox{8pt}{+}
\rule{1mm}{0mm}
\begin{picture}(52,32)
\put(0,0){\line(1,0){50}}
\put(0,0){\line(0,1){30}}
\put(50,0){\line(0,1){30}}
\put(0,30){\line(1,0){50}}
\put(10,15){\vector(1,0){30}}
\put(10,15){\circle*{3}}
\put(10,15){\circle{6}}
\put(40,15){\circle*{3}}
\end{picture}
\ \hfill\raisebox{8pt}{(\,11\,)}

\vskip8dd
\noindent
{\it Atom (neutral)
\rule{25mm}{0mm} Photon h$\nu$
\rule{4mm}{0mm} Atom (positive ionized)
\rule{17mm}{0mm} Electron e$^{-}$}
\vskip8dd

\noindent
This graph scheme (11) describes an atom influence on photon with
the ``destruction'' of the photon's h$\nu$~--- semicycle which as
if spontaneously is transformed into $e^{-}$ and $e^{+}$, with the
following inclusion of $e^{+}$ into the peripheral line of atom and
production of (+)-signed semiwalk \begin{picture}(30,20)
\put(0,0){\line(0,1){20}}
\put(0,20){\line(1,0){30}}
\put(0,0){\line(1,0){30}}
\put(30,0){\line(0,1){20}}
\put(0,10){\line(1,0){30}}
\put(22,0){\vector(-1,0){10}}
\put(7,10){\vector(1,0){10}}
\put(22,20){\vector(-1,0){10}}
\end{picture}
for positive
ionized atom.  The maximum kinetic energy, $E_{k}$, of the
photoelectron is given by Einstein's equation: $\displaystyle E_{k}
= h\nu - \varphi\,(\varphi$ is the work function of the solid). On
the other hand, graph equation (11) may be considered also as an
insertion of the photon's h$\nu$~--- semicycle into an ``atom~---
{\bf RvT}'' which is accompanied by the exchange emission back
again of an ``electron~--- {\bf RvT}''.

\subsubsection*{2.7.2 About so-called
atomic system ``fatigue''}

At last an efficient consideration of similar graph schemes---or
graph equations~--- for any typical microobject, having as a rule
the shell structure (with a presence of the sets of quantum
numbers), can be carried out only on condition that one taking into
account an obligatory ``isomerism'' in the complex {\bf RvT}~---
representation of every compound microobject~--- system and
additionally a many - value variety of every graph scheme with
participation of corresponding interacting reagents (photons,
electrons, etc.)

Obviously, that it is very naturally for a like situation to
suppose a possibility of the various failures in a complicated {\bf
RvT}~--- representation of the composite microsystems.

In this connection, one may arise with some probability a specific
shell system ``fatigue'', including  particularly a ``fatigue'' of
an atomic system with steady state sets, which can be appeared
perhaps in result of some default of a peripheral part of the
``graph isomers'', for example after an emission of photon and
suitable reconstruction of atom (see graph equation (9)), i. e. in
spite of an assertion of Ulam's hypothesis, or in result of some
transformation of this peripheral part into the balanced signed
semiwalk of excited atom already, in particular by an absorption of
photon (see graph equation (10)). Thus one may probably create an
accidental situation of non-full reproduction of atom (in spite of
an experimentally line spectra of excited atoms and ions with
enough strict invariability), connected straightly with many~-
``planes'' {\bf RvT}~--- representation of atomic system, which
will cause the combinatorial failure after a some number of the
like regular and even cyclic processes.

Analogous failure can be occured also at the photoelectric effect
in atom (see graph equation (11)) and generally in every graph
scheme with shell structure of the main ``fatigueless'' microobject
where figures the many~- ``planes'' {\bf RvT}~--- description.

\subsection*{2.8 Mass-classification of subnuclear microobjects\\
and ($r,\,s$)-subsequences}

Among the chaos of the experimental mass values $(M/m_{e})_{\rm
exp}$ [{\bf X}] (in $m_{e}$ unit, with $T_{v=2}=1$ for electron) of
``initial'' and ``excited'' subnuclear microobjects [{\bf X}] in
Table 2 of part II of paper (see Ref. [11]) may be traced the several
basic $(r,\,s)$~--- subsequences along with {\it a number of
``chief'' microobjects at} $r=s=0$ {\it corresponding to the pions}
$[{\bf \pi}^{0}]$ {\it and} $[{\bf \pi^{\pm}}]$ $(v=9,$
$\displaystyle T_{v=9}=286),$ {\it to the} $N$~--- {\it
baryons---just proton} $[{\bf p}]$ {\it and neutron} [{\bf n}]~---
{\it and some light unflavored and strange mesons} $(v=11,$
$\displaystyle T_{v=11}=1842),$ {\it to the charmed baryons}
$\displaystyle [{\bf \Lambda}^{+}_{c}],$ $\displaystyle [{\bf
\Sigma}_{c}],$ $\displaystyle [{\bf \Xi}^{+}_{c}],$ $\displaystyle
[{\bf \Xi}^{0}_{c}],$ $\displaystyle [{\bf \Omega}^{0}_{c}]$
{\it and a series of mesons and baryons} $(v=12,$ $\displaystyle
T_{v=12}=4766)$ and, probably, to the $d$~--- quark $(v=6,$
$\displaystyle T_{v=6}=20).$ Of course, the indicated ``chief''
microobjects at $r=s=0$ can be described completely by the
above-mentioned sets of ``hedgehog''~--- {\bf RvT}.

At first we have in Table 4 one of such $\displaystyle (r,\,s)$~---
subsequences, namely, {\it the main (0,2)-subsequence of lepton
masses, where two known leptons disposed at} $v=11$ $([\mu])$ {\it
and} $v=14$ $([{\bf \tau}]),$ and of meson and charmed meson
masses~--- at $v=13$ $([{\bf \rho}],$ $[{\bf \omega}])$ and $v=14$
([{\bf D}$^{\rm 0}$], [{\bf D}$^{\pm}$], [{\bf D}$^{\rm
\pm}_{s}$]).

{\tabcolsep=3dd
\begin{center}
\newcommand{\Sloppy}{\emergencystretch 3em \tolerance 9999
}
\let\sloppy=\Sloppy
\begin{tabular}{c|cccccccc}
\multicolumn{9}{c}{\bf Table~4. (0,2)-subsequence (``bi-triplets'')
of lepton and other particle masses}                           \\
\multicolumn{9}{c}{}                                      \\[-5dd]
\hline
\multicolumn{1}{c|}{}                                            &
\multicolumn{1}{c}{}                                             &
\multicolumn{1}{c}{}                                             &
\multicolumn{1}{c}{}                                             &
\multicolumn{1}{c}{}                                             &
\multicolumn{1}{c}{}                                             &
\multicolumn{1}{c}{}                                             &
\multicolumn{1}{c}{}                                             &
\multicolumn{1}{c}{}                                       \\[-5dd]
\multicolumn{1}{c|}{$v$}                                         &
\multicolumn{1}{c}{9}                                            &
\multicolumn{1}{c}{10}                                           &
\multicolumn{1}{c}{11}                                           &
\multicolumn{1}{c}{12}                                           &
\multicolumn{1}{c}{13}                                           &
\multicolumn{1}{c}{14}                                           &
\multicolumn{1}{c}{\ldots{}}                                     &
\multicolumn{1}{c}{18}                                           \\
\multicolumn{1}{c|}{}                                            &
\multicolumn{1}{c}{}                                             &
\multicolumn{1}{c}{}                                             &
\multicolumn{1}{c}{}                                             &
\multicolumn{1}{c}{}                                             &
\multicolumn{1}{c}{}                                             &
\multicolumn{1}{c}{}                                             &
\multicolumn{1}{c}{}                                             &
\multicolumn{1}{c}{}                                       \\[-5dd]
\hline
                                                                 &
                                                                 &
                                                                 &
                                                                 &
                                                                 &
                                                                 &
                                                                 &
                                                                 &
                                                           \\[-5dd]
$\displaystyle T_{v}/2^{r}3^{s}$                                 &
32                                                               &
80                                                               &
205                                                              &
530                                                              &
1387                                                             &
3664                                         &
&
191\,300                                                         \\
$\displaystyle (r=0,\,s=2)$                                      &
                                                                 &
                                                                 &
                                                                 &
                                                                 &
                                                                 &
                                                                 &
                                                                 &
                                                                 \\
                                                                 &
                                                                 &
                                                                 &
                                                                 &
                                                                 &
                                                                 &
                                                                 &
                                                                 &
                                                          \\[-10dd]
\hline
                                                                 &
                                                                 &
                                                                 &
                                                                 &
                                                                 &
                                                                 &
                                                                 &
                                                                 &
                                                          \\[-5dd]
both                                                             &
                                                                 &
                                                                 &
                                                                 &
                                                                 &
                                                                 &
                                                                 &
                                                                 &
                                                                 \\
leptons                                                          &
                                                                 &
                                                                 &
$[\mu]$                                                          &
                                                                 &
                                                                 &
$[\tau]$                                     &
&
                                                                 \\
                                                                 &
                                                                 &
                                                                 &
                                                                 &
                                                                 &
                                                                 &
                                                                 &
                                                                 &
                                                          \\[-5dd]
\hline
                                                                 &
                                                                 &
                                                                 &
                                                                 &
                                                                 &
                                                                 &
                                                                 &
                                                                 &
                                                          \\[-5dd]
other                                                            &
                                                                 &
                                                                 &
                                                                 &
                                                                 &
mesons                                                           &
charmed mesons                                                   &
                                                                 &
boson                                                            \\
particles                                                        &
                                                                 &
                                                                 &
                                                                 &
                                                                 &
$[\rho]$                                                         &
[{\bf D}$^{0}$], [{\bf D}$^{\pm}$]                               &
                                                                 &
[{\bf Z}$^{0}$]                                                  \\
                                                                 &
                                                                 &
                                                                 &
                                                                 &
                                                                 &
                                                                 &
charmed strange meson                                            &
                                                                 &
                                                                 \\
                                                                 &
                                                                 &
                                                                 &
                                                                 &
                                                                 &
$[\omega]$                                                       &
[{\bf D}$^{\pm}_{s}$]                                            &
                                                                 &
                                                                 \\
                                                                 &
                                                                 &
                                                                 &
                                                                 &
                                                                 &
                                                                 &
and oth.                                                         &
                                                                 &
                                                                 \\
                                                                 &
                                                                 &
                                                                 &
                                                                 &
                                                                 &
                                                                 &
                                                                 &
                                                                 &
                                                          \\[-5dd]
\hline
                                                                 &
                                                                 &
                                                                 &
                                                                 &
                                                                 &
                                                                 &
                                                                 &
                                                                 &
                                                          \\[-5dd]
                                                                 &
                                                                 &
                                                                 &
207                                                              &
                                                                 &
                                                                 &
3474                                                             &
                                                                 &
                                                                 \\
$\displaystyle \left(M/m_{e}\right)_{\rm exp}$ [{\bf X}]         &
\multicolumn{8}{l}{---\,---\,---\,---\,---\,---\,---\,---\,---\,
---\,---\,---\,---\,---\,---\,---\,---\,---\,---\,---\,---\,---\,
---\,---\,---\,---\,---\,---\,---}                \\
                                                                 &
                                                                 &
                                                                 &
                                                                 &
                                                                 &
1507--1530                                                       &
3649--3852                                                       &
                                                                 &
178\,448                                                         \\
                                                                 &
                                                                 &
                                                                 &
                                                                 &
                                                                 &
                                                                 &
                                                                 &
                                                                 &
                                                                 \\
\end{tabular}
\end{center} }
\vskip 6dd

The next main (1,0)-subsequence in Table~5 includes a whole
collection of various particles and consists from the masses of
five quarks [{\bf u}], [{\bf d}], [{\bf s}], [{\bf c}], [{\bf t}] at
$v=6,\,7,\,10,\,12,\,17,$ of strange mesons [{\bf K}$^{\pm}$], [{\bf
K}$^{0},$ $\displaystyle \bar{{\bf K}}^{0}$] at $v=11,$ of ${\bf
\Lambda^0}$-, ${\bf\Sigma}$- and ${\bf \Xi}$-baryons at $v=12$ and at last of
$c\bar{c}$-mesons $[\eta_{c}],$ [{\bf J}/${\bf \Psi}$],
$[\chi_{c}]$ at $v=13.$

\vskip 8dd
{\tabcolsep=3dd
\begin{center}
\newcommand{\Sloppy}{\emergencystretch 3em \tolerance 9999
}
\let\sloppy=\Sloppy
\begin{tabular}{c|ccccccccc}
\multicolumn{10}{c}{\bf Table~5. (1,0)-subsequence (``doublets'')
of quark and
particle masses}                                               \\
\multicolumn{10}{c}{}                                    \\[-5dd]
\hline
\multicolumn{1}{c|}{}                                            &
\multicolumn{1}{c}{}                                             &
\multicolumn{1}{c}{}                                             &
\multicolumn{1}{c}{}                                             &
\multicolumn{1}{c}{}                                             &
\multicolumn{1}{c}{}                                             &
\multicolumn{1}{c}{}                                             &
\multicolumn{1}{c}{}                                             &
\multicolumn{1}{c}{}                                             &
\multicolumn{1}{c}{}                                      \\[-5dd]
\multicolumn{1}{c|}{$v$}                                         &
\multicolumn{1}{c}{6}                                            &
\multicolumn{1}{c}{7}                                            &
\multicolumn{1}{c}{\ldots{}}                                     &
\multicolumn{1}{c}{10}                                           &
\multicolumn{1}{c}{11}                                           &
\multicolumn{1}{c}{12}                                           &
\multicolumn{1}{c}{13}                                           &
\multicolumn{1}{c}{\ldots{}}                                     &
\multicolumn{1}{c}{17}                                          \\
\multicolumn{1}{c|}{}                                            &
\multicolumn{1}{c}{}                                             &
\multicolumn{1}{c}{}                                             &
\multicolumn{1}{c}{}                                             &
\multicolumn{1}{c}{}                                             &
\multicolumn{1}{c}{}                                             &
\multicolumn{1}{c}{}                                             &
\multicolumn{1}{c}{}                                             &
\multicolumn{1}{c}{}                                             &
\multicolumn{1}{c}{}                                      \\[-5dd]
\hline
                                                                 &
                                                                 &
                                                                 &
                                                                 &
                                                                 &
                                                                 &
                                                                 &
                                                                 &
                                                                 &
                                                          \\[-5dd]
$\displaystyle T_{v}/2^{r}3^{s}$                                 &
10                                                               &
24                                                               &
                                                                 &
360                                                              &
921                                                              &
2383                                                             &
6243                                                             &
                                                                 &
317\,424                                                         \\
$\displaystyle (r=1,\,s=0)$                                      &
                                                                 &
                                                                 &
                                                                 &
                                                                 &
                                                                 &
                                                                 &
                                                                 &
                                                                 &
                                                                 \\
                                                                 &
                                                                 &
                                                                 &
                                                                 &
                                                                 &
                                                                 &
                                                                 &
                                                                 &
                                                                 &
                                                          \\[-5dd]
\hline
                                                                 &
                                                                 &
                                                                 &
                                                                 &
                                                                 &
                                                                 &
                                                                 &
                                                                 &
                                                                 &
                                                          \\[-5dd]
                                                                 &
                                                                 &
                                                                 &
                                                                 &
                                                                 &
meson                                                            &
baryons                                                          &
{\bf c\=c}-mesons                        &
                                                                 &
                                                                 \\
Hypothetical                                                     &
quark                                                            &
quark                                                            &
                                                                 &
quark                                                            &
$[\eta]$                                                         &
$\displaystyle [{\bf \Lambda^{\circ}}], [{\bf \Sigma}^{\pm}],$
$\displaystyle [{\bf \Sigma}^{0}]$                                                    &
$[\eta_{c}]$                                                     &
                                                                 &
quark                                                            \\
and real                                                         &
[{\bf u}]                                                        &
[{\bf d}]                                                        &
                                                                 &
[{\bf s}]                                                        &
strange                                                          &
$[{\bf \Xi}^{0}],$ $[{\bf \Xi}^{-}]$                             &
[{\bf J}/${\bf \Psi}]$                                           &
                                                                 &
[{\bf t}]                                                        \\
particles                                                        &
                                                                 &
                                                                 &
                                                                 &
                                                                 &
mesons                                                           &
quark                                                            &
                                                                 &
                                                                 &
                                                                 \\
                                                                 &
                                                                 &
                                                                 &
                                                                 &
                                                                 &
[{\bf K}$^{\pm}],$ [{\bf K}$^{0},$ $\bar{{\bf K}}^{0}]$          &
 [{\bf c}]                                                       &
$[\chi_{c}]$                                                     &
                                                                 &
                                                                 \\
                                                                 &
                                                                 &
                                                                 &
                                                                 &
                                                                 &
and oth.                                                         &
and oth.                                                         &
and oth.                                                         &
                                                                 &
                                                                 \\
                                                                 &
                                                                 &
                                                                 &
                                                                 &
                                                                 &
                                                                 &
                                                                 &
                                                                 &
                                                                 &
                                                         \\[-5dd]
\hline
                                                                 &
                                                                 &
                                                                 &
                                                                 &
                                                                 &
                                                                 &
                                                                 &
                                                                 &
                                                                 &
                                                          \\[-5dd]
                                                                 &
                                                                 &
                                                                 &
                                                                 &
                                                                 &
1071                                                             &
2183--2343                                                       &
                                                                 &
                                                                 &
                                                                 \\
$\displaystyle \left(M/m_{e}\right)_{\rm exp}$[{\bf X}]          &
10                                                               &
20                                                               &
                                                                 &
391                                                              &
                                                                 &
                                                                 &
5829--6870                                                       &
                                                                 &
                                                                 \\
                                                                 &
                                                                 &
                                                                 &
                                                                 &
                                                                 &
966--974                                                         &
2544--2585                                                       &
                                                                 &
                                                                 &
352\,250                                                         \\
                                                                 &
                                                                 &
                                                                 &
                                                                 &
                                                                 &
                                                                 &
                                                                 &
                                                                 &
                                                                 &
                                                                 \\
\end{tabular}
\end{center} }
\vskip 6dd

Below we add two more Tables~6 and 7 with (2,0)- and
(2,1)-subsequences of the particle masses which contain the baryon
$\displaystyle [{\bf \Omega}^{-}]$ at $v=13$ and the quarks
[{\bf c}] and [{\bf b}] at $v=14,$ and also mentioned already quark
[{\bf s}] at $v=12.$

\vskip 8dd
{\tabcolsep=3dd
\begin{center}
\newcommand{\Sloppy}{\emergencystretch 3em \tolerance 9999
}
\let\sloppy=\Sloppy
\begin{tabular}{c|cccc}
\multicolumn{5}{c}{\bf Table~6. (2,0)-subsequence
(``bi-doublets'') of quark and
particle masses}                                               \\
\multicolumn{5}{c}{}                                      \\[-5dd]
\hline
\multicolumn{1}{c|}{}                                            &
\multicolumn{1}{c}{}                                             &
\multicolumn{1}{c}{}                                             &
\multicolumn{1}{c}{}                                             &
\multicolumn{1}{c}{}                                      \\[-5dd]
\multicolumn{1}{c|}{$v$}                                         &
\multicolumn{1}{c}{13}                                           &
\multicolumn{1}{c}{14}                                           &
\multicolumn{1}{c}{\ldots{}}                                     &
\multicolumn{1}{c}{17}                                           \\
\multicolumn{1}{c|}{}                                            &
\multicolumn{1}{c}{}                                             &
\multicolumn{1}{c}{}                                             &
\multicolumn{1}{c}{}                                             &
\multicolumn{1}{c}{}                                      \\[-5dd]
\hline
                                                                 &
                                                                 &
                                                                 &
                                                                 &
                                                          \\[-5dd]
$\displaystyle T_{v}/2^{r}3^{s}$                                 &
3122                                                             &
8243                                                             &
                                                                 &
158\,712                                                         \\
$(r=2,\,s=0)$                                                    &
                                                                 &
                                                                 &
                                                                 &
                                                                 \\
                                                                 &
                                                                 &
                                                                 &
                                                                 &
                                                          \\[-5dd]
\hline
                                                                 &
                                                                 &
                                                                 &
                                                                 &
                                                          \\[-5dd]
Hypothetical                                                     &
baryon                                                           &
quark                                                            &
                                                                 &
boson                                                            \\
and real                                                         &
$[{\bf \Omega}^{-}]$                                             &
[{\bf b}]                                                        &
                                                                 &
[{\bf W}$^{\pm}$]                                                \\
particles                                                        &
and oth.                                                         &
and oth.                                                         &
                                                                 &
                                                                 \\
                                                                 &
                                                                 &
                                                                 &
                                                                 &
                                                         \\[-5dd]
\hline
                                                                 &
                                                                 &
                                                                 &
                                                                 &
                                                         \\[-5dd]
$\displaystyle \left(M/m_{e}\right)_{\rm exp}$ [{\bf X}]         &
3273                                                             &
8415                                                             &
                                                                 &
157\,201                                                       \\
\end{tabular}
\end{center} }
\vskip 4dd
\vskip 6dd
{\tabcolsep=3dd
\begin{center}
\newcommand{\Sloppy}{\emergencystretch 3em \tolerance 9999
}
\let\sloppy=\Sloppy
\begin{tabular}{c|ccccc}
\multicolumn{6}{c}{\bf Table~7. (2,1)-subsequence of quark and
particle masses}                                                \\
\multicolumn{6}{c}{}                                      \\[-5dd]
\hline
\multicolumn{1}{c|}{}                                            &
\multicolumn{1}{c}{}                                             &
\multicolumn{1}{c}{}                                             &
\multicolumn{1}{c}{}                                             &
\multicolumn{1}{c}{}                                             &
\multicolumn{1}{c}{}                                      \\[-5dd]
\multicolumn{1}{c|}{$v$}                                         &
\multicolumn{1}{c}{12}                                           &
\multicolumn{1}{c}{13}                                           &
\multicolumn{1}{c}{14}                                           &
\multicolumn{1}{c}{15}                                           &
\multicolumn{1}{c}{16}                                           \\
\multicolumn{1}{c|}{}                                            &
\multicolumn{1}{c}{}                                             &
\multicolumn{1}{c}{}                                             &
\multicolumn{1}{c}{}                                             &
\multicolumn{1}{c}{}                                             &
\multicolumn{1}{c}{}                                      \\[-5dd]
\hline
                                                                 &
                                                                 &
                                                                 &
                                                                 &
                                                                 &
                                                          \\[-5dd]
$\displaystyle T_{v}/2^{r}3^{s}$                                 &
397                                                              &
1040                                                             &
2748                                                             &
7318                                                             &
19\,615                                                          \\
$(r=2,\,s=1)$                                                    &
                                                                 &
                                                                 &
                                                                 &
                                                                 &
                                                                 \\
                                                                 &
                                                                 &
                                                                 &
                                                                 &
                                                                 &
                                                          \\[-5dd]
\hline
                                                                 &
                                                                 &
                                                                 &
                                                                 &
                                                                 &
                                                          \\[-5dd]
                                                                 &
                                                                 &
meson                                                            &
quark                                                            &
``excited''                                                      &
``initial''                                                      \\
Hypothetical                                                     &
quark                                                            &
$[\eta]$                                                         &
[{\bf c}]                                                        &
{\bf c}$\bar{{\bf c}}$-mesons                                   &
and                                                              \\
and real                                                         &
[{\bf s}]                                                        &
strange                                                          &
baryons                                                          &
                                                                 &
``excited''                                                      \\
particles                                                        &
                                                                 &
mesons                                                           &
$[{\bf \Xi}^{0}],$ $[{\bf \Xi}^{-}]$                             &
                                                                 &
{\bf b}$\bar{{\bf b}}$-mesons                                   \\
                                                                 &
                                                                 &
[{\bf K}$^{\pm}$], [{\bf K}$^{0},$ $\bar{{\bf K}}^{0}]$          &
and oth.                                                         &
                                                                 &
                                                                 \\
                                                                 &
                                                                 &
                                                                 &
                                                                 &
                                                                 &
                                                         \\[-5dd]
\hline
                                                                 &
                                                                 &
                                                                 &
                                                                 &
                                                                 &
                                                          \\[-5dd]
                                                                 &
                                                                 &
1071                                                             &
2544                                                             &
                                                                 &
18\,513--                                                        \\
$\displaystyle \left(M/m_{e}\right)_{\rm exp}$ [{\bf X}]         &
391                                                              &
                                                                 &
                                                                 &
6683--7906                                                       &
--21\,564                                                        \\
                                                                 &
                                                                 &
966--974                                                         &
2573--2585                                                       &
                                                                 &
                                                                 \\
                                                                 &
                                                                 &
                                                                 &
                                                                 &
                                                                 &
                                                                 \\
\end{tabular}
\end{center} }
\vskip 6dd

It will be noted that baryon $[{\bf \Lambda}^{0}]$ belongs also to
(1,1)-subsequence at $v=13$ whereas in (0,1)-subsequence must be
included charmed strange mesons [{\bf D}$^{\pm}_{s}$], [{\bf
D}$^{*\pm}_{s}$] and charmed baryon $[{\bf \Lambda}^{+}_{c}]$ at
$v=13$ and also bottom mesons [{\bf B}$^{\pm}$], [{\bf B}$^{0}$],
[{\bf B}$^{*}$], bottom strange meson [{\bf B}$^{0}_{s}$] and
bottom baryon $[{\bf \Lambda}^{0}_{b}]$ at $v=14.$

These particular varieties of $(r,\,s)$-subsequences to a certain
degree remind the group combinations~--- ``conglomerates'' or
multiplets of particles from a more sophisticated ``numerology'' in
the framework of special well-known formalism based on the
symmetry principle with a common interpretation. In our case a
notion of ``multiplet of particles'' arises from a natural analysis
simply of ``$\frac{1}{2}$-part'' or ``$\frac{1}{3}$-part'' of the
``whole'', for example $T_{v},$ and is regulated alternatively by
the hierarchical rules of a double- and a triple-splitting
mechanism for the ``hedgehogs''~--- {\bf RvT}s on a base of the
evident recurrent inequalities for $T_{v}$ (see particularly in
section 4 of part~II of paper~--- Ref. [11]). Thus the
``doubleness'' (``doubletness'') as well as the ``tripleness''
(``tripletness'') are simply a result of use of the
``hedgehog''~--- {\bf RvT} formalism for a representation of the
discrete microobjects as the fundamental constituents of all the
matter in the universe.

Under any realization of this ``doubletness''---``tripletness''
graph concept one must set up the various possibilities and lastly
some definite rules for a more exact practical estimation of mass
parameters of the discrete microobjects along with a rough approach
to $T_{v}$ and its fractions $T_{v}/2^{r}3^{s}$ in part~II of paper
(see Ref. [11]).

At first, starting from the recurrent inequalities for $T_{v}$ in a
form
$$\displaystyle \frac{T_{v+1}}{2}\geqslant
T_{v}>\frac{T_{v+1}}{3}\eqno(12)                                 $$

\noindent
one can determine an ``interval-function'' $\Delta\,(T_{v})$ which
allows to calculate for every $v$ the distribution range of
$T_{v}$-values (see below Table~8).

\vskip 8dd
{\tabcolsep=9dd
\begin{center}
\newcommand{\Sloppy}{\emergencystretch 3em \tolerance 9999
}
\let\sloppy=\Sloppy
\begin{tabular}{c|c|c||c|c|c}
\multicolumn{6}{c}{\bf Table~8. ``Interval-function''
$\bf \Delta\,(T_{v})=\frac{1}{6}T_{v+1}$}                     \\
\multicolumn{6}{c}{}                                      \\[-5dd]
\hline
\multicolumn{1}{c|}{}                                            &
\multicolumn{1}{|c|}{}                                           &
\multicolumn{1}{|c||}{}                                           &
\multicolumn{1}{c|}{}                                           &
\multicolumn{1}{|c|}{}                                           &
\multicolumn{1}{|c}{}                                     \\[-5dd]
\multicolumn{1}{c|}{$v$}                                         &
\multicolumn{1}{|c|}{$T_{v}$}                                    &
\multicolumn{1}{|c||}{$\Delta\,(T_{v+1})$}                        &
\multicolumn{1}{c|}{$v$}                                        &
\multicolumn{1}{|c|}{$T_{v}$}                                    &
\multicolumn{1}{|c}{$\Delta\,(T_{v+1})$}                         \\
\multicolumn{1}{c|}{}                                            &
\multicolumn{1}{|c|}{}                                           &
\multicolumn{1}{|c||}{}                                           &
\multicolumn{1}{c|}{}                                           &
\multicolumn{1}{|c|}{}                                           &
\multicolumn{1}{|c}{}                                     \\[-5dd]
\hline
                                                                 &
                                                                 &
                                                                 &
                                                                 &
                                                                 &
                                                          \\[-5dd]
3                                                                &
2                                                                &
0,6(6)                                                           &
9                                                                &
286                                                              &
119,83(3)                                                        \\
4                                                                &
4                                                                &
1,5                                                              &
10                                                               &
719                                                              &
307                                                              \\
5                                                                &
9                                                                &
3,3(3)                                                           &
11                                                               &
1842                                                             &
794,3(3)                                                         \\
6                                                                &
20                                                               &
8                                                                &
12                                                               &
4766                                                             &
2081                                                             \\
7                                                                &
48                                                               &
19,16(6)                                                         &
13                                                               &
12\,486                                                          &
5495,5                                                           \\
8                                                                &
115                                                              &
47,6(6)                                                          &
14                                                               &
32\,973                                                          &
14635,16(6)
\end{tabular}
\end{center} }
\vskip 6dd

Thus in Tables~4--7 we have an essentially original
``concentration'' of the whole sets of particle group masses, with
``mixing'' of meson and baryon particle categories, near the
definite number of vertices (especially within an interval {\it
v}=11$\div$14) of corresponding ``hedgehogs''~--- {\bf RvT}~s which
in various configurations are responsible for the structural
peculiarity of different microobjects and in such a way one can be
set up their concrete physical characteristics. In this connection
one must elaborate for the developing graph formalism an adequate
graph interpretation which could allow to reproduce some general
physical picture. At that one can be taken into consideration:
a)~all above-stated graph (``hedgehog''~--- {\bf RvT})
regularities, pointed out as different ({\it r, s})~---
subsequences for microobject masses; b)~the further graph schemes
(see, for example, equations (9)--(11)) for various atomic and
nuclear reactions with conservation as well as non-conservation of
a number of vertices and, moreover, c)~also the developing group
formations for other microobject characteristics, including
``charge'', ``spin'' and so on, that are in agreement with the
studying graph formalism.

Table 8 indicates a sharp increase of $\Delta$ ({\it
T$_{v}$})---values with {\it v}, in other words an enlargement of
the possibility diapason for the ``hedgehog''~--- {\bf RvT}
formalism in respect to a description of mass parameters as well as
of other characteristics of microobjects including the heavy
physical and further the complicated biological microobjects.

\section*{3 TWO-LAYER PHYSICS ACCORDING\\
TO HEISENBERG~--- DYSON}

It still remains a mystery why the physicists were not among
the first investigators of a graph aspect of Kirchhoff's work about
an electric network of 1847 (see Ref. [22]) and are enough
indifferent so far to the development of graph~--- or, more
exactly, tree~--- theory which now is a section of the discrete
mathematics wide-using in applied sciences. Apparently, such
passivity occurs as a result of the full hegemony of the continuous
mathematics in physics and correspondingly the preferable
construction of the so-called one-``plane'' continuous models,
in opposition to the initially many-``planes'' discrete models how
it is easy to see from the above-pointed examples of the discrete
mathematics application. Therefore a break-through in
``discontinuity'' didn't happen before in physics.

\pagebreak
\subsection*{3.1 Peculiarities of microobjects description \\
in Heisenberg---Dyson's physics}

Up till now through a development of the continuous physical
models, based on the methods of corresponding sections of a
continuous mathematics, the most part of an appropriate concept
notions can be included in a contemporary theoretical
representation of natural events and can be used as a base for an
introducing any new model. However after the transition to
microworld many conceptual conclusions, following from the pointed
continuous models, have turned out experimentally proofless and even
senseless that is due to some earlier included concept notions
which haven't adequate interpretation.

Among such non-adequate notions in old concept one may call
artificial quasi-objects like the quarks, although similar
quasi-objects can be predicted as the certain ({\it r, s})~---
subsequences in the framework of a proposed graph formalism for
discrete microobjects in a ``hedgehog''~--- {\bf RvT}
representation (see section~2). Indeed, in Tables 5--7 one comes to
light a ``multiplet'' from five quarks in (1,0)~--- subsequence:
{\it T$_{v}$}/2 at {\it v=6}  ([{\bf u]}~--- ``up''), {\it v=7}
([{\bf d}]~--- ``down''),  {\it v=10} ([{\bf s}]~--- ``strange''),
{\it v=12} ([{\bf c}]~--- ``charm''),
{\it v=17} ([{\bf t}]~--- ``top'') and an alternative ``doublet''
from (2, 1)~--- subsequence: {\it T$_{v}$}/12 at
{\it v=12} ([{\bf s}]~---
``strange'') and {\it v}=14 ([{\bf c}]~--- ``charm'').  Besides
these main subsequences for the fundamental constituents in an old
concept one must mention in passing some other kind of ``doublet''
from the quarks at one value of {\it v}=14:[{\bf c}]~---
``charm'' from (2,1)~--- subsequence ({\it T$_{v}$}/12) and [{\bf
b}]~--- ``beauty'' or ``bottom'' from (2,0)~--- subsequence ({\it
T$_{v}$}/4).

In this arisen consideration it is noteworthy also the creation of
a new ``top hierarchy'' of particles with a distinct separation of
the ``chief'' particles, including the pions and the {\it N}~---
baryons (at {\it r=s=0}), and the main leptons in (0,2)~---
subsequence, whereas any other categories of particles may be
interpreted as the following, possibly ``derivative'', layers of
such ``top hierarchy''. Here it is important to
emphasize also that in quantum
mechanics one can set up only a peculiar discontinuity of dynamical
quantities~--- depending from the sets of quantum numbers~--- for
physical microobjects but we haven't yet a straight quantum
description of their discrete complicated structure and, still
more, of their many-``planes'' representation beyond space---time.

Meanwhile an existing problem of description of the microobjects
structure and their interactions beyond space---time is solved above
(in section~2) in the framework of linear two-layer aproach or
two-layer matrix (1) (at $\displaystyle \delta \equiv \alpha$)
approximation by means of the specially introduced notions of the
``graph geometry'' in a linear ``hedgehog''~--- {\bf RvT}
representation and the linear loop-forming quasi-continuous
field ``object'' as an oriented multigraph with some loops (composed
linearly from two or more ``hedgehogs''~--- {\bf RvT} s responsible
for separate discrete microobjects). Just in this way one can
occur the linear two-layer physics of microworld according to
Heisenberg---Dyson with an inclusion of the new essential features
of atomic and nuclear stable systems and subnuclear particles.

\subsection*{3.2 Macro networks in Heisenberg---Dyson's physics}

On the other hand, Heisenberg---Dyson's physics of macroworld
coincides with Kirchhoff---Maxwell's two-layer physics approach,
formulated in part I of paper (see Ref. [10]) and constructed on
the two-layer matrix (1) at $\displaystyle \delta \equiv \omega$
with skeleton trees basis. By using of such skeleton trees basis
one can write down the all possible solutions for various macro
networks, as a some graph prototype of any macroobject. In the
main, the linear equations, derived from two-layer matrix
representation, allow to set up the linear structure of all
fragments of any macro network i. e. the linear superposition of
corresponding real sets of {\bf SvT}~s, determined by famous
Kirchhoff's theorem (see for example, Ref. [12]).

In this connection, including the total of subsection 3.1, it is
immensely important to note that {\it on the whole a graph (tree)
conception}, appeared initially from the linear algebraic
Kirchhoff's laws (Ref. [22]) in the form of {\bf SvT}~---
formalism, may be considered generally {\it as an one-value result
of the ``universal linearity'' of physical laws of the nature}.
Although, according to Ulam's reminiscence, yet Fermi on this
occasion said: ``Well, it does not say in the Bible that
fundamental equations of physics must be linear''.

According to an originated problem of the creation of an initial
pre-condition for adequate graph description of macroworld it will
be noted that any physical theory for the consideration of
different discrete or quasi-continuous macroobjects can be
constructed on a base of the concrete simplest ``geometrical
figures'' (in a general sense) taking into account the main
peculiarities of these natural macroobjects and giving a ``native''
representation of their structure in the framework of existed
models picture. These ``geometrical figures'', in other words, can
be applied as a some physical image of the complex macro process
with a participation of macroobjects and, lastly, these ones must
be transformed to the corresponding equivalent graphs (trees)
according to some macroworld rules, in agreement with the
Kirchhoff's rules and the corresponding macroworld principles. In
this way we should have a generation of the graph prototype for
macroobject as an analog of macro network (compare, for example,
the various ``geometrical figures'' for crystals, macromolecules,
biological objects, etc.).

It is remarkably in a development of the continuous physical theory
for macroworld that already at the stage of an appearance of the
Maxwell's equations for electromagnetic field, written down here in
two-layer matrix form
$$\displaystyle\left(
\begin{array}{c}
-\displaystyle\vec{\bf D}\raisebox{11pt}{$\!\!\!\!\bullet$}
+{\rm curl}\,\vec{\bf H}
\\ ic\, {\rm div}\,{\bf D} \end{array} \right )=\left(
\begin{array}{c}
\vec{\bf j} \\
ic\,\rho
\end{array}
\right), \quad
\left(
\begin{array}{c}
\displaystyle
\vec{\bf B}\raisebox{11pt}{$\!\!\!\!\bullet$}
+{\rm curl}\,\vec{\bf E} \\
c\, {\rm div}\,\vec{\bf B} \end{array}
\right)=\left(
\begin{array}{c}
\vec{\theta} \\
0
\end{array}
\right), \ \eqno (13) $$

\noindent
one was carried out, with a a great virtuosity, the crossing over
of the components of ``charge'' part
$\displaystyle \left(\begin{array}{c}
\vec{\bf j}\\
ic \rho \end{array}\right)$
(a known 4-vector) and the components of remaining ``curl field''
part with $\vec{\bf E}$, $\vec{\bf D}$, $\vec{\bf H}$,
$\vec{\bf B}$.  As a result one could obtain then, without any
transition to the discrete physical objects conception, only a
``mixed'' picture that don't allow to separate directly the upper
layer (discrete material objects) and the under layer
(quasi-continuous field ``objects'') in the framework of a
proposed Heisenberg---Dyson's two-layer matrix (1) scheme, right
however for the solutions of pointed equations (13). And, at last,
starting from the Maxwell's equations in emptiness, one should
mention the description of electromagnetic radiation where the
energy can be regarded as waves propagated through usual space
requiring no supporting medium: in this case figures only one under
layer.

\pagebreak
\subsection*{3.3 Heisenberg~--- Dyson's two-layer approach to study \\
of microobjects structure}

Returning again to microworld in the course of an analysis of the
Heisenberg~--- Dyson's two-layer physics one must take active part
firstly in the furhter decoding of {\it the postulated ``physical
graph'' as a some derivative}, already within discrete two-layer
matrix scheme, {\it analog of a solution of the differential
equations for continuous physical models}. In other words, {\it
while passing through a stage of the physical dynamics i.~e. a
stage of an investigation of corresponding differential equation
itself} (with Newton's, Lagrange's, ``evolution's'' and other
differential operators) {\it in favour of a two-layer matrix
approximation for its solution, one can go over to a
straightforward explanation of the structure and the properties of
discrete physical microobjects}, without reference to dynamics
aspects in space---time. Substantially, a two-layer matrix
approximation for pointed solution must be derived by the way of
its extremal transformation to the two-layer system of linear
algebraic equations (for a natural description of ``physical
graph'', as an analog of  ``Kirchhoff's laws graph'').

After {\bf RvT}~--- estimation of the subnuclear microobject masses
(an average error ranges 0 to 10 per cent) in part II of paper (see
Ref. [11]) we obtain some rough mass-classification with the
``concentration''  of all microobject masses within an interval of
{\bf RvT}~- vertices: $v=9\div 14$, as an illustration of the
enough adequacy of ``hedgehog''~--- {\bf RvT} formalism to the
clearing of microworld picture. Nevertheless, at the more accurate
and complete determination of the structure and the properties of
discrete microobjects the leading role must play their interactions
(or interaction ``charges'' according to Table~3) and connected
with them different ``core'' characteristics, including ${\frak
M}^{\rm c}$, $v^{\rm c}_{F}$, {\bf C}$_{v_{F}}$, as
the main representants of hierarchical sets of ``hedgehog''~--- {\bf
RvT}.

However, formerly it is important to elucidate from the point of
view of a ``graph language'' (as an alternative ``third language'')
that the dynamical quantity, namely
``{\it E}~--- energy'' provided for the continuous models (side by
side with ``{\it L}~--- Lagrangian'', ``{\it H}~--- Hamiltonian'',
etc.) must be ``measured'' already with the help of dimensionally
identical parameter ``$c^{2}$'' ([$E$]$\sim$[$c^{2}$]) giving a
purely numerical~--- dimensionless~--- result for mass $m=E/c^{2}$.
Indeed, according to the special theory  of relativity the mass of
microobject is a measure of its total energy content that is
equivalent in a ``graph language'' to the dimensionless $T_{v}$~---
numbers or their fractional ($r, s$)~--- ``fragments''
$T_{v}$/2$^{r}$3$^{s}$ and is, analogously, in a some agreement
with the known relation of two periods $\omega=2\pi/T$ in a
``wave language'', i.~e. the mathematical period 2$\pi$ and the
physical period $T$. Therefore, in the case of subnuclear particles
there are a sequence of integers $T_{v}$ and fractions
$T_{v}$/2$^{r}$3$^{s}$ for the straightforward determination of
their masses using namely a ``native'' $T_{v}$~- quantity
of ``hedgehog''~--- {\bf RvT} formalism.

Further, according to our consideration in subsection 2.6, it is
significant also that the nature of any microobject mass cann't be
determined by one type of interaction inasmuch as for any microobject
there are simultaneously all types of interaction (four or more)
which determine conjointly a general (``mixed'') mass of every
microobject and, consequently, of any macroobject.

\pagebreak
\subsubsection*{3.3.1 Responsibility of different types of
interaction  \\ for complicated microobjects structure}

A ``hedgehog''~--- {\bf RvT} representation for a single pure
``core'' of nucleons (n$^{\circ}$, p$^{\rm +}$) at $v=11$, i.~e.
{\bf C}$_{v_{F}=10}$ with maximum ``free'' vertices
$v^{c}_{F}=v^{\rm max}_{F}$ in Table~3, corresponds apparently to a
some set of free constituents~--- quarks which are responsible for
an inner ``invisible'' structure of the strong interacting nucleons
and, according to the quark---confinement hypothesis, cannot escape
from one another (in QCD, particularly, interactions between
quarks get weaker or stronger as the space---time distance between
them gets smaller or greater, correspondingly). This unwarranted
assumption about a role of the single pure ``core'' {\bf
C}$_{v_{F}=10}$ for the strong interaction of a nucleon stipulates
an introduction of the additional suggestion that one of the
``planes'' in the ``hedgehog''~--- {\bf RvT} representation of a
nucleon can never be ``seen'', even in isolation, and at that may
be shown only by a some indirect way. Such suggestion concides with
a conclusion from section 2.6 about the ``microworld situation''
within which there is an ``invisible ``core''-vertices
confinement'' for any typical discrete microobject. It is easy to
see, however, that a presence of the ``charges'' altogether
and quantum numbers in the single pure ``core'' of
nucleons {\bf C}$_{v_{F}=10}$ cannot respond to an integer of the
free constituents~--- quarks, i.~e. in our case a linear set of the
various quarks in nucleons in the form of superposition sum may be
considered simply as a special combinatorial factor for an
estimation of the real composition of their ``core''-vertices (as
opposed to compensation rules for the ``charge'' multiplets with a
concrete, for example, fractional electric charge assigned to the
different quarks).

The next single one-``plane'' {\bf RvT}~--- representation or
``hedgehog''~--- {\bf RvT} with $v_{F}=v^{\rm max}_{F}-1=9, \
v^{\rm c}_{F}=8$, which is responsible for the weak interaction of
nucleon in Table~3, on the contrary is quite ``visible'' and takes
part in the decays (for example, $n^{\circ}$ in Table~3) and in the
construction of the definite real physical systems (for example,
nuclei from the sets of separate $n^{\circ}$ and $p^{+}$). Indeed
in following Table 9 one depict the most evident typical graph
constructions of some light nuclei from the separate
``hedgehogs''~--- {\bf RvT}s of neutrons and protons, within a
general ``core'' with several root vertices.
Over this ``core'' one must arrange the sets of
steady states, or one can potentially join the hierarchical schemes
of corresponding electronic shell levels controlled by
electromagnetic~--- or other long-ranged~--- interaction, with 1840
``hedgehogs''~--- {\bf RvT}s for every nucleon (according to
Table~3).

\vskip 14dd

\begin{center}
{\bf Table~9. ``Core''~--- structure of some light atomic nuclei}

\end{center}

\vskip 4dd

{\tabcolsep=3dd
\newcommand{\Sloppy}{\emergencystretch 3em \tolerance 9999
}
\let\sloppy=\Sloppy

\begin{center}
\noindent
\begin{tabular}{c|c|c}
\hline
\multicolumn{1}{c|}{}                                    &
\multicolumn{1}{|c|}{}                                   &
\multicolumn{1}{|c}{}                                   \\
\multicolumn{1}{c|}{Nucleus of}                          &
\multicolumn{1}{|c|}{Nucleus of}                         &
\multicolumn{1}{|c}{Nucleus of}                         \\
\multicolumn{1}{c|}{hydrogen atom}                       &
\multicolumn{1}{|c|}{deuterium atom}                     &
\multicolumn{1}{|c}{helium atom}                        \\ [5dd]
\hline
\multicolumn{1}{c|}{\rule{25mm}{0mm}}                    &
\multicolumn{1}{|c|}{\rule{42mm}{0mm}}                   &
\multicolumn{1}{c}{\rule{90mm}{0mm}}                    \\
\multicolumn{1}{c|}{
\begin{picture}(50,50)
\put(0,5){\line(0,1){40}}
\put(0,45){\line(1,0){50}}
\put(0,5){\line(1,0){50}}
\put(50,5){\line(0,1){40}}
\put(20,40){\line(0,-1){20}}
\put(20,40){\circle*{3}}
\put(20,40){\circle{6}}
\put(20,40){\line(-1,-2){10}}
\put(20,40){\line(1,-1){20}}
\put(20,20){\circle*{3}}
\put(10,20){\circle*{3}}
\put(40,20){\circle*{3}}
\put(40,0){\vector(0,1){20}}
\put(40,0){\circle*{3}}
\put(25,20){\circle*{1}}
\put(30,20){\circle*{1}}
\put(35,20){\circle*{1}}
\put(18,9){(p$^+$)}
\put(32,-10){(e$^+$)}
\end{picture}
}&
\multicolumn{1}{|c|}{
\begin{picture}(105,50)
\put(0,5){\line(0,1){40}}
\put(0,45){\line(1,0){100}}
\put(0,5){\line(1,0){100}}
\put(100,5){\line(0,1){40}}
\put(20,40){\line(0,-1){20}}
\put(20,40){\circle*{3}}
\put(20,40){\circle{6}}
\put(20,40){\line(-1,-2){10}}
\put(20,40){\line(1,-1){20}}
\put(20,20){\circle*{3}}
\put(10,20){\circle*{3}}
\put(40,20){\circle*{3}}
\put(10,0){\line(0,6){20}}
\put(10,0){\circle*{3}}
\put(25,20){\circle*{1}}
\put(30,20){\circle*{1}}
\put(35,20){\circle*{1}}
\put(22,9){(n$^\circ$)}
\put(50,0){
\put(20,40){\line(0,-1){20}}
\put(20,40){\circle*{3}}
\put(20,40){\circle{6}}
\put(20,40){\line(-1,-2){10}}
\put(20,40){\line(1,-1){20}}
\put(20,20){\circle*{3}}
\put(10,20){\circle*{3}}
\put(40,20){\circle*{3}}
\put(40,0){\vector(0,1){20}}
\put(40,0){\circle*{3}}
\put(25,20){\circle*{1}}
\put(30,20){\circle*{1}}
\put(35,20){\circle*{1}}
\put(18,9){(p$^+$)}
\put(32,-10){(e$^+$)}
}
\end{picture}
}&
\multicolumn{1}{|c}{
\begin{picture}(205,50)
\put(0,5){\line(0,1){40}}
\put(0,45){\line(1,0){200}}
\put(0,5){\line(1,0){200}}
\put(200,5){\line(0,1){40}}
\put(20,40){\line(0,-1){20}}
\put(20,40){\circle*{3}}
\put(20,40){\circle{6}}
\put(20,40){\line(-1,-2){10}}
\put(20,40){\line(1,-1){20}}
\put(20,20){\circle*{3}}
\put(10,20){\circle*{3}}
\put(40,20){\circle*{3}}
\put(10,0){\line(0,6){20}}
\put(10,0){\circle*{3}}
\put(25,20){\circle*{1}}
\put(30,20){\circle*{1}}
\put(35,20){\circle*{1}}
\put(22,9){(n$^\circ$)}
\put(50,0){
\put(20,40){\line(0,-1){20}}
\put(20,40){\circle*{3}}
\put(20,40){\circle{6}}
\put(20,40){\line(-1,-2){10}}
\put(20,40){\line(1,-1){20}}
\put(20,20){\circle*{3}}
\put(10,20){\circle*{3}}
\put(40,20){\circle*{3}}
\put(10,0){\vector(0,1){20}}
\put(10,0){\circle*{3}}
\put(25,20){\circle*{1}}
\put(30,20){\circle*{1}}
\put(35,20){\circle*{1}}
\put(20,9){(p$^+$)}
\put(2,-10){(e$^+$)}
}
\put(100,0){
\put(20,40){\line(0,-1){20}}
\put(20,40){\circle*{3}}
\put(20,40){\circle{6}}
\put(20,40){\line(-1,-2){10}}
\put(20,40){\line(1,-1){20}}
\put(20,20){\circle*{3}}
\put(10,20){\circle*{3}}
\put(40,20){\circle*{3}}
\put(10,0){\vector(0,1){20}}
\put(10,0){\circle*{3}}
\put(25,20){\circle*{1}}
\put(30,20){\circle*{1}}
\put(35,20){\circle*{1}}
\put(20,9){(p$^+$)}
\put(2,-10){(e$^+$)}
}
\put(150,0){
\put(20,40){\line(0,-1){20}}
\put(20,40){\circle*{3}}
\put(20,40){\circle{6}}
\put(20,40){\line(-1,-2){10}}
\put(20,40){\line(1,-1){20}}
\put(20,20){\circle*{3}}
\put(10,20){\circle*{3}}
\put(40,20){\circle*{3}}
\put(40,0){\line(0,1){20}}
\put(40,0){\circle*{3}}
\put(25,20){\circle*{1}}
\put(30,20){\circle*{1}}
\put(35,20){\circle*{1}}
\put(22,9){(n$^\circ$)}
}
\end{picture}
}\\[25dd]
\end{tabular}
\end{center}

\vskip 14dd

It is already known that positive ``charge'' of nucleus {\bf Z} is
determined by a number of separate in-side directed lines beyond
``core'' connecting with a number of included into ``core''
protons. These in-side directed lines emerge in the result of a
dissolution of lines in some oriented multigraphs (with semicycles for adequate
single non-directed lines) which must be disposed beyond the
corresponding ``core'' and may be considered as if the ``free
radicals'' of nuclei with interacting ``charges'' (compare,
analogously, the ``free radicals'' into the like presentation of
chemical or biological molecules). Obviously that additionally to
Table~9, a ``core'' of any next, more heavy, nuclei with {\bf
Z}$>$2 can be immediately constructed in a similar manner.

It will be noted also that a like form of the ``core''~---
structure of atomic nuclei, especially in the region of enough
large {\bf Z}, presupposes a some possible ``fission'' process
similar to the biological cell ``fission'' according to the
principle of replication as an evolution of structures with several
hierarchical centres.

\subsubsection*{3.3.2 ``Fission'' and ``fusion'' of ``core''---vertices under
formations \\
and structural changes of microobjects}

Every type of interaction between various microobjects as a cause
of their structural changes may be realized by means of exchange
forces resulting from the continued interchange of ``transferable
particles'' (photons, mesons, gluons, intermediate bosons, etc.) in
a manner that bonds their hosts together. For example, it is
suggested that in the strong interaction gluons are exchanged
between quarks or mesons are exchanged between nucleons; at that
mesons jumped from proton to neutron and back again. Also in a case
of the electromagnetic interaction there is the covalent bond
involving electrons and so on.

On the other hand, all changes in the structure of every real,
certainly complicated, microobject, participating in different
physical processes (decays, collisions, scattering, productions,
reactions, etc.), are accompanied with a ``fission'' and a
``fusion'' of ``core''~--- vertices, including the corresponding
root vertices, that may be interpreted as a reconstruction, a
reproduction, a replication (or a multiplication), and so on, of
``core'' for varions {\bf RvT}s.

For concrete illustration from the very beginning it is expedient
to consider the simplest example of complex nucleon structure with
the set of {\bf RvT}s at $v=11$ in which one must embed all kinds
of microobject with the sets of {\bf RvT}s at $v\leqslant 11$,
namely in the form of superposition sum of Riemann's ``counting
homogeneous elements''  for altogether constituent parts of the
single pure ``core'' of nucleons ($n^{\circ}, \ p^{+}$), i.~e.
${\bf C}_{v_{F}=10}\equiv [{\bf N}](T_{v=11}$):
$$[{\bf N}]
(T_{v=11})=k_{1}\cdot[\eta](T_{v=11}/2)+k_{2}\cdot [{\bf
K}](T_{v=11}/2)+k_{3}\cdot[\mu](T_{v=11}/9)+k_{4}\cdot[{\bf
s}](T_{v=10}/2)+k_{5}\cdot\{[\pi](T_{v=9})+$$
$$+[\pi](T_{v=10}/3)\}+
k_{6}\cdot\{[{\bf d}](T_{v=6})+[{\bf d}](T_{v=7}/2\}+
k_{7}\cdot [{\bf u}](T_{v=6}/2)+\sum\limits^{4}_{i=1}k_{7+i}\cdot
[{\bf X}](T_{v=6-i})+\cdots{},\eqno (14)$$

\noindent
where the last member contains the
pure ``core'' {\bf RvT}s with ${{\frak M}}^{\rm c}_{\rm max}=2(v-1)$
for 2$\leq v \leq$5 as ``graph vacancies'' {\bf C}$_{v_{F}=i}$;
$i$=1,2,3,4 including $e^{\pm}$, various ${\rm \nu}$ and others.
Here we use the next designations

\parbox[c]{140mm}{
\vskip 8dd
{\tabcolsep=6dd
\begin{center}
\begin{tabular}{l|lll}
$[{\bf N}]$-nucleons:
$[{\bf n]},[{\bf p}]$
& $v=11$ & $e=0,+1$ & $T_{v=11}=1842$\\ &&&\\[-8dd] \hline
&&&\\[-8dd]
$[{\bf u}]$-quark                                                  &
$v=6$ &
$e=+2/3$ &
$T_{v=6}/2=10$\\
$[{\bf d}]$-quark &
$v=6,7$ &
$e=-1/3$ &
$T_{v=6}=20, \ T_{v=7}/2=24$ \\
$[\pi]$-mesons:
$[\pi^{\pm}],[\pi^{0}]$ &
$v=9,10$ &
$e=\pm1,0$ &
$T_{v=9}=286, \
T_{v=10}/3=240$ \\
$[{\bf s}]$-quark &
$v=10$ &
$e=-1/3$ &
$T_{v=10}/2=360$ \\
$[\eta]$-meson &
$v=11$ &
$e=0$ &
$T_{v=11}/2=921$ \hfill $\lefteqn{\qquad\ \ (15)}$\\
$[{\bf K}]$-mesons:$[{\bf K}^{\pm}]$, $[{\bf
K^{0}}{\bf \bar{K}^{0}}]$ &
$v=11$ &
$e=\pm1,0$ &
$T_{v=11}/2=921$ \\
$[\mu]$-leptons: $[\mu^{\pm}]$ &
$v=11$ &
$e=\pm 1$ &
$T_{w=11}/9=205.$
\\ \end{tabular} 
\end{center} }
\vskip 8dd }



Nevertheless, a problem of the Riemann's ``counting homogeneous
elements'' in expression (14)--(15) can be solved only in the frame
of evident discrete {\bf RvT}~--- transformations embraced the
necessary permutations of various ``core''-vertices, after
admissible operations of their ``fission'' and ``fusion'' and
``core''-generating or ``core''-eliminating processes, and also
specially the ``invisible ``core''-vertices confinement'' within
the nucleon for ``free-walking'' quark vertices (without root
vertex). The last remark concerns two coefficients $k_{6}$ and
$k_{7}$ in (14).

It will be noted also that a former Yukawa's suggestion about the
exchange forces between nucleons that held them together involved
several generating pions which now, in expression (14), are simply
included as a graph constituent part of nucleon with the
maintenance of the principles of a graph integrity of nucleon and
separately of pions (and any other particles), of a graph
non-reducing of nucleon as a whole to the sum of parts and, at
last, of a graph reproduction of nucleon.

The further study of the more complicated microobjects may be
carried out analogously.

In general terms, under a penetration into the structure and
a formation of the theoretical model characteristics of concrete
microobject it is necessary to leave out its dynamics and
everyone from artificial symmetry multiplets, determined solely the
non-structural quasi-discontinuity of ``conditional microobject''
in a form of the discrete sets of quantum numbers~--- so-called
``QN-discontinuity'' (with the following increase of these sets for
the classification aims and an experimentally proof comparison).
Just then one must essentially alter former description starting
from an introduction of the true discontinuity beyond space---time
(instead of ``QN-discontinuity'') and a direct calculation of the
precise proton mass as an initial point of the ``top hierarchy''
consideration. As it is demonstrated already such true
discontinuity beyond space~--- time can arise only by an acceptance
of the axiom of the universal linearity of physical laws in
macroworld as well as in microworld; at that above-mentioned ``top
hierarchy'' for microobjects is a natural consequence of the same
linearity. Indeed, an initial principle of superposition in the
framework of graph formalism may be fulfiled in the form of a sum
of the Riemann's ``counting homogeneous elements'' including the
``hedgehogs''~--- {\bf RvT}s, the {\bf SvT}s and the other graph
``planes'' of discrete physical objects, correspondingly.

     In the end,  summing-up specially the main results of an attained
two-layer  physics  approach  to microworld it is easy to see that the
set of {\bf RvT}s determines the peculiarities of the following many
- "planes"  representations of real single microobjects and
complex microob- ject - systems:

          -one strong interaction "core" (one set of pure "core"
vertices) which is responsible probably for the notion of
"Elementary Particle" or for some microobject corresponding only
to the set of "invisible "core" vertices",

          -one weak interaction "core" (one set of "core"
vertices with separate vertex beyond "core") which is responsible
for the decaying microoobjects - "neutral" as well as "charged",
having the single "visible" vertices beyond "core",

          -several electromagnetic or gravitational interaction
"cores" (several sets of various "core" vertices with additional
vertices beyond these "cores") which are responsible altogether
for different shell microsystems.

     Using the far-seeing propositions of Ulam's hypothesis for
graphs (see Ref.[12]), proved later for trees (Refs.[14-16]),
it will be noted that every tree (including everyone {\bf RvT}) can
be completely presented as a some composition of respective
subtrees with the same or less number of vertices. Then any pure
"core" for microobject may be written down in the form of
superposition sum of some set of pure "cores" for its constituent
parts.  Therefore, particularly, every "Elementary Particle" can be
decomposed as a "whole" in the superposition sum of Riemann's
"counting homogeneous elements" as suitable separate "parts"
(perhaps such transitions from one "invisible" "core" to the
postulating Riemann's sum of suitable several "invisible" "cores"
are typically for the strong interacting microobjects whereas if
we consider the weak interaction one must turn to an analysis of
the decaying microobjects and "visible" "charge" reactions). On
the other hand, this decomposition must reflect simultaneously a
hierarchical non-reducing of the "whole" to the set of "parts"
that may be connected with the discretion of {\bf RvT}-basis
of"graph geometry" beyond space-time as opposed to the usual bases
in "external" space-time with "point-continuity" of coordinate axes
(Cartesian frame of reference, for example) which, however,
deprive of personal character of discrete physical microobject in
corresponding theoretical models.

\section*{4 Conclusions}

The proposed model of an introduction of the discrete physical
objects in micro- as well as in macrophysics is realized by means of
postulating of the new ``physical graph'' as a discrete microobject
and simultaneously using the ``Kirchhoff's laws graph'' for an
electric ``macro network''. These graphs may be presented through
the discrete sets of root {\it v}-trees ({\bf RvT}~---
``hedgehog'') for microobjects and skeleton {\it v}-trees ({\bf
SvT}) for any typical macroobject i.~e. beyond space---time,
inasmuch as the pointed sets of trees form in some way the {\bf
RvT}-basis and the {\bf SvT}-basis, correspondingly. In this case
the common ``Extension axiom'' in a space---time is replaced by the
more adequate ``Binding axiom'' beyond space---time. The rejection
from space---time agrees, on the other hand, with Eddington's
supposition (see Ref. [23]) that the space and the time are only
the approximate notions which must be substituted finally by the
more general idea of an arrangement of any events of nature with
adapting of the more adequate mathematical formalism and getting
accustomed to an absence of generally accepted space---time.

If we are at the some stage where one should realize a discrete
object scheme for the description of physical microworld or, more
generally, for the creation of an unified physical theory of micro- as
well as macroworld then again one may appear a ``choice situation''
in part of a branching of the directions of development of physical
theory (see Ref.~[9]). This ``choice situation'' is analogous to a
similar one in the fundamental physical theories in the mid-1960s
(see, particularly, a narration about competition of the {\it
S}-matrix theory, the quantum field theory and the group theory
in Ref.~[8]) or earlier (Riemann, 1868; P\'oincare, 1905; and so
on). In fact, a discrete formulation of the fundamental physical
theory, which should consider, in particular, as a variety of the
discrete mathematics, didn't realize during those periods.

Now one may formulate the principal results of a given paper
(parts~I--III), which are got in the framework of discrete graph
formalism for Heisenberg---Dyson's two-layer physics, in
comparison with the theses of an old and today's continuous theory.

1. For the first time in the late-1950s one had believed (see, in
particular, Landau papers Refs.~[13, 2]) that solely a direct use of
non-perturbative diagram technique is completely consistent and
could serve as a ground for an adequate successive construction of
the future fundamental theory without an adaptation to any former
continuous theory (QED, QCD and so on) or physical models.
Factually it was an initial attempt to show the way for an
inevitable introduction of the discontinuity in the present
physical theory in a form of special diagram technique or, in terms
of this paper, of graph kinematics of discrete physical objects.

2. The most significant and original conclusion following from the
graph kinematics, in the frame of the Heisenberg---Dyson's
two-layer approach (analysed in this part~III), consists in a
many-``planes'' {\bf RvT} or {\bf SvT} representation of any
discrete physical object in opposition to one-``plane'' physical
objects in continuous physical models. Essentially, that every
property and every type of interaction of a concrete discrete
microobject can be connected with the definite subsets of such
``planes'' from the full set of {\bf RvT}~--- representation
``planes'' for a given microobject. In any single case these
separate ``planes'' or their sets must be experimentally shown,
directly or indirectly, for every discrete physical object as if
crystal faces, for example.

3. A many-``planes'' representation of concrete discrete object in
a form of the ``hedgehog''~--- {\bf RvT} or {\bf SvT} sets may be
originated by passing through a stage of the physical dynamics in
space~--- time and by postulating beyond space~--- time either a
``physical graph'' as a some extremally derivative algebraic analog
of the solution of differential equations for continuous
microobject models or simply a ``Kirchhoff's laws graph'' as a
ground for the construction of a solution of already algebraic
equations for macro network. Using the Heisenberg---Dyson's
two-layer scheme one can now practically go over to the two-layer
matrix approximation on the base (1) which composed from an
incidence {\bf I} and a loop {\bf CD}$\displaystyle (\delta)$ graph
matrices.

4. Next very important and unexpected result concerns to the
superposition of different types of interaction or simultaneous
presence of the four (or more) sorts of interaction ``charges'' all
round for any microobject (see Table~3). In the framework of
many-``planes'' ``hedgehog''~--- {\bf RvT} representation the
short-ranged interactions (weak, strong) can be described by an
one-``plane'' ``hedgehog''~--- {\bf RvT} with maximum values of
$\displaystyle {\frak M}^{c}$
reflected a ``compact'' inner
structure of any microobject in respect to the given interactions.
Quite the contrary, for the long-ranged interactions
(gravitational, electromagnetic) there are the sets of
many-``planes'' ``hedgehog''~--- {\bf RvT} with ${\frak
M}^{c}<{\frak M}^{c}_{\rm max} - 1$ determined an enough
``transparent'' inner structure of the same microobject which may
be considered as an adequate hierarchical scheme of levels or
``orbitals'' (see Tables~2--3).  Now a central problem of the
theory shifts to the decoding and interpretation of a
many-``charge'' discrete microobject in the frame of two-layer
Heisenberg---Dyson's ``hedgehog''~--- {\bf RvT} formalism.

5. Based on the incidence matrix {\bf I} ``graph geometry'' for the
real discrete physical objects describes all main characteristics
reflected their peculiar many-``planes'' inner structure in the
frame of purely discrete mathematics formalism beyond common
space~--- time. There is an essential problem to perform the
crucial experiment for a suitable determination of these
characteristics.

6. The notion of such kind as an interacting ``charge'' of
microobject in the graph kinematics concept may be extracted only
by means of the loop matrix {\bf CD}$\displaystyle (\alpha)$ from
the symbolical quantities for the loop-forming quasi-continuous
field ``objects'' in Heisenberg---Dyson's two-layer matrix
approximation.

7. Strong correlation between upper (discrete material objects) and
under (quasi-continuous field ``objects'') layers was set up with
the help of Maxwell's equations (13) and as a reassured result
{\it the ``mixed'' non-linear picture of nature} with the
components of ``charge'' 4-vector $\left({\vec{\bf
j}\atop{ic\rho}}\right)$ and the components of ``curl fields''
$\displaystyle \vec{\bf E}$,
$\displaystyle \vec{\bf D}$,
$\displaystyle \vec{\bf H}$,
$\displaystyle   \vec{\bf B}$
{\it may be eliminated simply by
the way of an introduction of the discrete physical objects
conception} in the framework of two-layer matrix approximation
for the linear Heisenberg---Dyson's scheme. This example
illustrates also {\it a general conclusion about the ``universal
linearity'' beyond space---time.}

8. In contrast to quantum mechanics, the set up earlier problems of
a stability (or reproduction) of definite stationary states, of a
transition itself (or ``jump'') of electron between stationary
states and of a sudden emission (or ``production'') of discrete
photon in atoms may be considered naturally and straightforwardly
on the base of Heisenberg---Dyson's two-layer graph kinematics for
discrete atomic shell systems beyond space~--- time. The
simultaneous or ``parallel'' participation of all ``planes'' (or
stationary forms) of any stable shell system in ``hedgehog''~---
{\bf RvT}'s description means its permanent reproduction by
non-open ``transferability'' between ``graph isomers'' with the
same number of vertices.

9. It is quite possible, that there are different combinatorial
failures in a complicated {\bf RvT}~--- representation of the
many-``planes'' discrete models for composite shell microsystems as
opposed to one-``plane'' continuous models. It this way one may
emerge, in particular, a ``fatigue'' of an atomic shell system with
steady state sets (in spite of strictly line spectra of excited
atoms), which can be interpreted as a result of any default of a
peripheral part of the ``graph isomers'' (some breaking
reproduction of initial atom) or as a result of various
transformations of this peripheral part into the balanced signed
semiwalk of excited atom.

10. In the frame of Heisenberg---Dyson's two-layer physics one
may introduce the ``true discontinuity'' straightly for an
initially structural discrete physical microobject on a ``graph
language'' beyond space---time, instead of ``QN~---
discontinuity'' (the discrete sets of various free~--- introduced
quantum numbers for the non-structural quasi-continuous
``conditional microobject'' and its dynamical quantities in
space---time). In this way one can present the precise nucleon
mass, as an initial point of the ``top hierarchy'' consideration,
by an acceptance of the axiom of universal linearity of physical
laws. At that the principle of superposition must be realized in
the form of a sum of Riemann's ``counting homogeneous elements''
including a particular variant of the set of ``hedgehogs''~--- {\bf
RvT}s for constituent parts of nucleon as a whole.

11. For the certain setting up of an adequacy of ``hedgehog''~---
{\bf RvT} approach to the clearing of microworld picture one must
carry out at least a rough classification for a total list of the
experimental values of subnuclear microobject masses using the
fractional ``fragments'' 2$^{-r}$3$^{-s}$ of the root trees numbers
$T_{v}$. At average error of such $T_{v}$-estimation, ranged 0 to
10 per cent, there take place the ``concentration'' of all masses
within an interval of {\bf RvT}-vertices $v=9\div14$ (see part~II)
and the ``tracing'' of the several basic ($r, s$)~--- subsequences
for lepton, quark and other real and hypothetical particle masses
(see Tables 4--7).

12. With the use of natural ``hedgehog''~--- {\bf RvT}
regularities, as the first iteration in a making of different ($r,
s)$-subsequences for microobject masses, one can derive a new
``top hierarchy'' of particles, instead of more sophisticated
``numerology'' in the frame of artificial symmetry multiplets
within an old ``QN~--- discontinuity'' concept. This more regular
``top hierarchy'' possesses a feature of the distinct separation of
the masses of ``chief'' particles, including electron
($T_{v=2}$=1), pions ($T_{v=9}$=286), $N$~--- baryons and some
mesons ($T_{v=11}$=1842) in basic (0,~0)-subsequence (initial 1-st
level), and also the masses of main leptons $\mu$ and $\tau$ in
(0,~2)-subsequence (3-rd level).  Any other category of particles
belongs to the next more deep levels of such ``top hierarchy'' with
the fixed ``chief'' particles ($e, \ \pi, \ N$) and several mesons
and charmed baryons.

13. The ``doubleness'' 2$^{-r}$ (or ``mass doubletness''
$T_{v}/2^{r}$) as well as the ``tripleness'' $3^{-s}$ (or ``mass
tripletness'' $T_{v}/3^{s}$) for subnuclear microobject masses are
simply a natural consequence of the ``hedgehog''~--- {\bf RvT}
formalism application to the specific representation of these
discrete microobjects as fundamental constituents of the matter.
Along with a rough approach to subnuclear microobject masses by
means of $T_{v}$ and its fractions $T_{v}/2^{r}3^{s}$, reflected
permanently ``$\frac{1}{2}$~--- parts'' or ``$\frac{1}{3}$~---
parts'' of the ``whole'', one may develop this ``doubletness''~---
``tripletness'' {\bf RvT} concept further for a more exact
determination of these masses using some other discrete microobject
characteristics including various ``charges'', ``spins'', and so
on, for specification of such ``top hierarchy''.

14. The exchange forces between protons, that held them together,
involve usually several generating pions. Now in the framework of
Heisenberg---Dyson's approach to graph kinematics of the discrete
microobjects these pions are simply included (without any
generation) as the autonomic graph constituent parts of proton. In
a given form of description one must maintain: (1)~the principle of
a graph integrity for {\bf RvT}s of proton (number of vertices
$\displaystyle v$=11) and separately for {\bf RvT}s of pions
(number of vertices $\displaystyle v$=9) and also for {\bf RvT}s of
any other constituent parts (number of vertices
$\displaystyle v$\,$\displaystyle \leqslant$\,11) of proton;
(2)~the principle of a graph non-reducing for {\bf RvT}s of proton
as a whole to the sum for {\bf RvT}s of autonomic constituent
parts; (3)~the principle of a graph reproduction (an intermediate
mechanism) for {\bf RvT}s of proton through {\bf RvT}s of pions.

15. Leptons $\displaystyle \mu$ and $\displaystyle \tau$ may be
considered as the possible original excited states of lepton {\it
e} with an increased number of vertices:
$$\displaystyle v_{\mu} = v_{e} + \Delta v_{\mu} = 2 + 9 = 11;
\quad v_{\tau} = v_{e} + \Delta v_{\tau} = 2 + 12 = 14.          $$

\noindent
In particular, a muon should be looked as an excited electron for
which $\displaystyle v_{\mu}$ concides with $\displaystyle v_{p} =
v_{n} = 11$ and $\displaystyle \Delta v_{\mu} = v_{\pi} = 9.$

\end{document}